\documentclass[]{aa} 
\usepackage{graphicx}
\usepackage{epsfig}
\usepackage{natbib}
\bibpunct{(}{)}{;}{a}{}{,}

\begin{document}


\titlerunning{The CORALIE survey for southern extra-solar planets. XII}

\title{The CORALIE survey for southern extra-solar planets.}
\subtitle{XII. Orbital solutions for 16 extra-solar planets discovered
  with CORALIE.  \thanks{Based on observations collected with the
    {\footnotesize CORALIE} echelle spectrograph on the 1.2-m Euler
    Swiss telescope at La\,Silla Observatory, ESO Chile} \thanks{The
    precise radial velocities presented in this paper are available in
    electronic form at the {\footnotesize CDS} via anonymous ftp to
    {\tt cdsarc.u-strasbg.fr (130.79.128.5)} or via {\tt
      http://cdsweb.u-strasbg.fr/cgi-bin/qcat?J/A+A/}, except for the
    multi-planet systems that will appear in a future paper describing
    their dynamical evolutions, taking planet-planet interaction into
    account} }

\author{M.~Mayor\inst{1}
   \and S.~Udry\inst{1}  
   \and D.~Naef\inst{1} 
   \and F.~Pepe\inst{1}
   \and D.~Queloz\inst{1} 
   \and N.C.~Santos\inst{1,2}
   \and M. Burnet\inst{1}
}

\offprints{michel.mayor@obs.unige.ch}

\institute{Observatoire de Gen\`eve, 51 ch. des Maillettes, CH-1290
  Sauverny, Switzerland 
\and Centro de Astronomia e Astrof{\'\i}sica da Universidade de Lisboa,
Observat\'orio Astron\'omico de Lisboa, Tapada da Ajuda, 1349-018
Lisboa, Portugal }

\date{Received 02-09-2003/ Accepted 30-09-2003-IN-PRESS-IN-PRESS-IN-PRESS } 

\abstract{This paper summarizes the information gathered for 16 still
  unpublished exoplanet candidates discovered with the CORALIE echelle
  spectrograph mounted on the Euler Swiss telescope at La Silla
  Observatory. Amongst these new candidates, 10 are typical extrasolar
  Jupiter-like planets on intermediate- or long-period
  (100\,$<$\,$P$\,$\leq$\,1350\,d) and fairly eccentric
  (0.2\,$\leq$\,$e$\,$\leq$\,0.5) orbits (HD\,19994, HD\,65216,
  HD\,92788, HD\,111232, HD\,114386, HD\,142415, HD\,147513,
  HD\,196050, HD\,216437, HD\,216770).  Two of these stars are in
  binary systems. The next 3 candidates are shorter-period planets
  (HD\,6434, HD\,121504) with lower eccentricities among which we find
  a hot Jupiter (HD\,83443).  More interesting cases are given by the
  multiple-planet systems HD\,82943 and HD\,169830. The former is a
  resonant $P_2/P_1$\,=\,2/1 system in which planet-planet
  interactions are influencing the system evolution. The latter is
  more hierarchically structured.
 
}

\maketitle

\section{Introduction}

For more than 5 years the {\footnotesize CORALIE} planet-search
programme in the southern hemisphere \citep{Udry-2000:a} has been
ongoing at the 1.2-m Euler Swiss telescope -- designed, built and
operated by the Geneva Observatory -- at La Silla Observatory
(ESO/Chile).  During these 5 years, {\footnotesize CORALIE} has
allowed the detection (or has contributed to the detection) of 38
extra-solar planet candidates. This substantial contribution together
with discoveries from various other programmes have provided a sample
of more than 115 exoplanets that now permits us to point out
interesting statistical constraints for the planet formation and
evolution scenarios \citep[see e.g.][ for reviews on different aspects
of the orbital-element distributions or primary star
properties]{Mayor-2003,Udry-2003:a,Santos-2003:a}.

The majority of our {\footnotesize CORALIE} exoplanet candidates have
been published in a series of dedicated papers\footnote{Another
  dedicated series has also been started for close binaries requiring
  a 2-dimensional correlation analysis for radial-velocity estimates.
  The method has already revealed a 2.5-M$_{\rm Jup}$ planet orbiting
  the primary \citep{Zucker-2003:b} and a 19-M$_{\rm Jup}$ brown dwarf
  orbiting the secondary \citep{Zucker-2003:a} of the HD\,41004 close
  visual binary system}, the latest among them reporting the detection
of the shortest-period Hot Jupiter discovered by radial-velocity
surveys around {\footnotesize HD}\,73256 \citep{Udry-2003:b} and the
very interesting case of {\footnotesize HD}\,10647
\citep{Udry-2003:c}, a star with a high IR excess indicative of the
presence of a debris disk.  The present paper of this series describes
the {\footnotesize CORALIE} exoplanets that have not been published
yet.  This subsample includes candidates announced several months ago,
rapidly after their detection to allow follow-up observations. It also
includes some candidates with very long periods or that are members of
multi-planet systems requiring a delay in their final analysis. Also,
some of the new candidates correspond to very recent detections.

\begin{table*}[t!]
\caption{
\label{table1}
Observed and inferred stellar parameters for the 
stars hosting planets presented in this paper. Definitions and
sources of the quoted values are given in the text. The age and
rotational period estimates are based on calibrations of the 
$R^\prime_{HK}$ activity indicator \citep{Donahue-93,Noyes-84}, 
whose reference source is also indicated: (S) for this paper following 
\citet{Santos-2000:b}, (H) for \citet{Henry-96} and (B) for 
\citet{Butler-2002}.  The applied analyses and uncertainty estimates 
can be found in the quoted references.}
\begin{tabular}{lllccccccrccllr}
\hline
HD       &Sp     &V &$B-V$  &$\pi$  &$M_V$ &$L$ 
&$T_{\rm eff}$   &$\log{g}$   &\multicolumn{1}{l}{\hspace*{-1mm}$[Fe/H]$}  
&$M_\star$  &$v\sin i$  
&\hspace*{-2mm}$\log(R^{\prime}_{HK})$ &age
&\multicolumn{1}{l}{\hspace*{-2mm}$P_{\rm rot}$}\\ 
 & & & &[mas] & &[L$_{\odot}]$ &[$^\circ$\,K] &[cgs] & &[M$_{\odot}$]
 &[km/s] & &[Gy] &\hspace*{-2mm}[day] \\
\hline
6434         &G3IV  &\hspace*{-1mm}7.72 &0.613 &24.80 &4.69 &1.12
    &5835    &4.60  &\hspace*{-1mm}$-0.52$ &0.79   &2.3
    &\hspace*{-2mm}$-4.89$ (H) &3.8   &\hspace*{-2mm}18.6 \\           
19994        &F8V   &\hspace*{-1mm}5.07 &0.575  &44.69  &3.32   &3.81
    &6217    &4.29  &\hspace*{-1mm}0.25    &1.34   &8.1
    &\hspace*{-2mm}$-4.77$ (S) &2.4   &\hspace*{-2mm}12.2 \\
65216        &G5V   &\hspace*{-1mm}7.97 &0.672  &28.10  &5.21   &0.71
    &5666    &4.53  &\hspace*{-1mm}$-0.12$ &0.92   &$<$\,1
    &\multicolumn{1}{c}{\hspace*{-2mm}--}  &\multicolumn{1}{c}{--} 
    &\multicolumn{1}{c}{\hspace*{-2mm}--}  \\
82943        &G0    &\hspace*{-1mm}6.54 &0.623  &36.42  &4.35   &1.50
    &6005    &4.45  &\hspace*{-1mm}0.32 &1.15   &1.7
    &\hspace*{-2mm}$-4.82$ (S) &2.9   &\hspace*{-2mm}18.0   \\
83443        &K0V   &\hspace*{-1mm}8.23 &0.811  &22.97  &5.04   &0.88
    &5454    &4.33  &\hspace*{-1mm}0.35 &0.90   &1.4
    &\hspace*{-2mm}$-4.85$ (B) &3.2   &\hspace*{-2mm}35.3 \\
92788        &G5    &\hspace*{-1mm}7.31 &0.694  &30.94  &4.76   &1.05
    &5821    &4.45  &\hspace*{-1mm}0.32 &1.10   &1.8
    &\hspace*{-2mm}$-4.73$ (S) &2.1   &\hspace*{-2mm}21.3 \\
111232       &G5V   &\hspace*{-1mm}7.59 &0.701  &34.63  &5.29   &0.69
    &5494    &4.50  &\hspace*{-1mm}$-0.36$ &0.78   &1.2
    &\hspace*{-2mm}$-4.98$ (H) &5.2  &\hspace*{-2mm}30.7 \\ 
114386       &K3V   &\hspace*{-1mm}8.73 &0.982  &35.66  &6.49   &0.29
    &4804    &4.36  &\hspace*{-1mm}$-0.08$    &0.68   &1.0
    &\multicolumn{1}{c}{\hspace*{-2mm}--}  &\multicolumn{1}{c}{--} 
    &\multicolumn{1}{c}{\hspace*{-2mm}--}\\ 
121504       &G2V   &\hspace*{-1mm}7.54 &0.593  &22.54  &4.30   &1.55
    &6075    &4.64  &\hspace*{-1mm}0.16 &1.18   &2.6
    &\hspace*{-2mm}$-4.57$ (S) &1.2   &\hspace*{-2mm}8.6  \\ 
142415       &G1V   &\hspace*{-1mm}7.33 &0.621  &28.93  &4.64   &1.14
    &6045    &4.53  &\hspace*{-1mm}0.21 &1.03   &3.3
    &\hspace*{-2mm}$-4.55$ (S) &1.1  &\hspace*{-2mm}9.6  \\ 
147513       &G3/5V &\hspace*{-1mm}5.37 &0.625  &77.69  &4.82   &0.98
    &5883    &4.51  &\hspace*{-1mm}0.06 &1.11   &1.5
    &\hspace*{-2mm}$-4.38$ (S) &0.3   &\hspace*{-2mm}4.7  \\ 
169830       &F8V   &\hspace*{-1mm}5.90 &0.517  &27.53  &3.10   &4.59
    &6299    &4.10  &\hspace*{-1mm}0.21    &1.40   &3.3
    &\hspace*{-2mm}$-4.82$ (S) &2.8   &\hspace*{-2mm}8.3 \\ 
196050       &G3V   &\hspace*{-1mm}7.50 &0.667  &21.31  &4.14   &1.83
    &5918    &4.34  &\hspace*{-1mm}0.22    &1.10   &3.1 
    &\hspace*{-2mm}$-4.65$ (S) &1.6   &\hspace*{-2mm}16.0 \\
216437      &G4IV/V &\hspace*{-1mm}6.04 &0.660  &37.71  &3.92   &2.25
    &5887    &4.30  &\hspace*{-1mm}0.25    &1.06   &2.5
    &\hspace*{-2mm}$-5.01$ (H) &5.8   &\hspace*{-2mm}26.7 \\
216770       &K0V   &\hspace*{-1mm}8.11 &0.821  &26.39  &5.22   &0.79
    &--      &--    &\hspace*{-1mm}0.23    &0.90  &1.4
    &\hspace*{-2mm}$-4.84$ (H) &3.1   &\hspace*{-2mm}35.6 \\ 
\hline
\end{tabular}
\end{table*}

The paper is organized as follows. In the next section we summarize
the primary star properties. The radial-velocity measurements and
inferred orbital solutions will be presented in Sect.\,3. In the last
section we summarize the results and provide some concluding remarks.

\begin{figure*}[th!]
\begin{center} 
\psfig{width=0.75\hsize,file=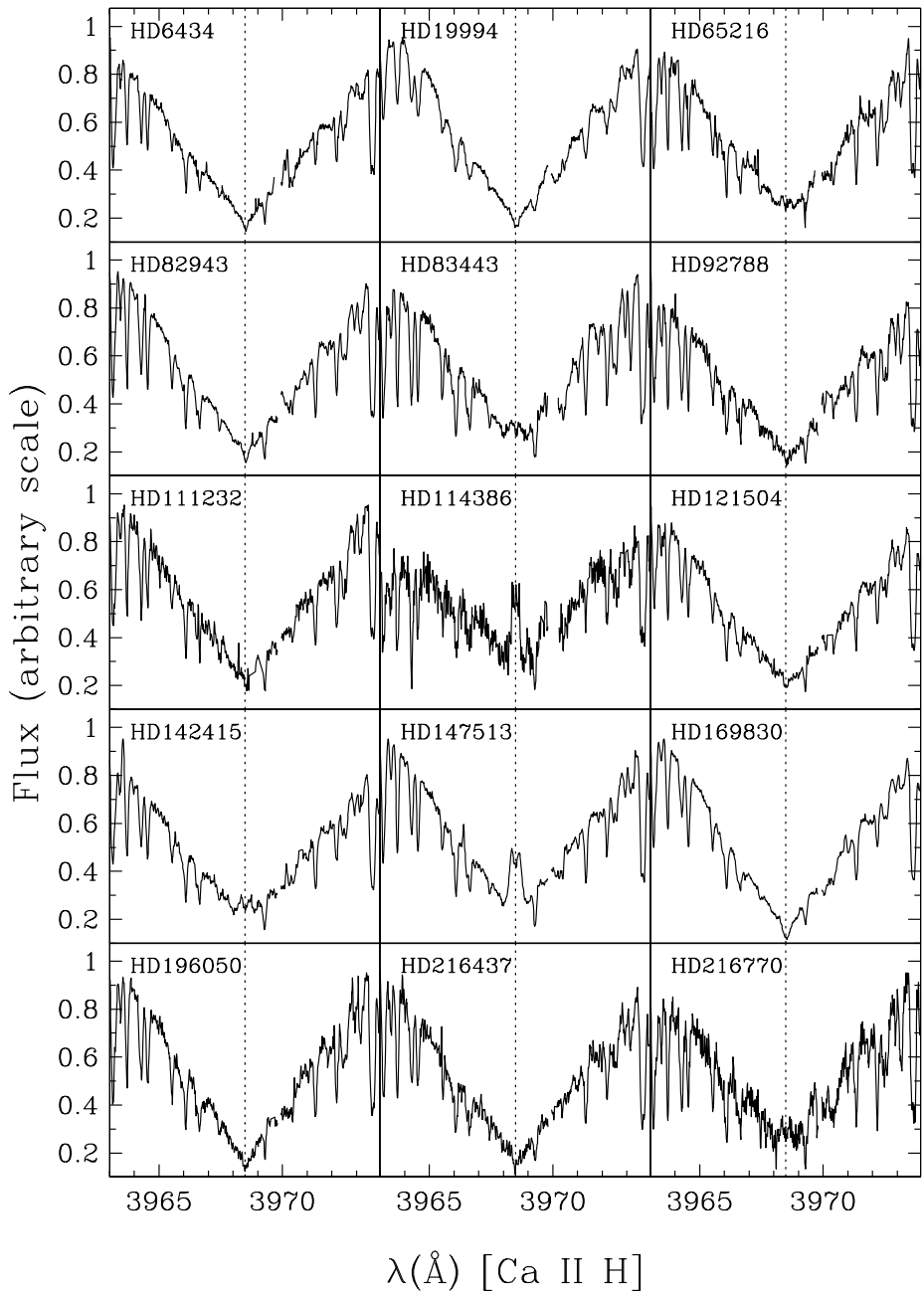}
\end{center}
\caption{
\label{fig1}
\mbox{$\lambda$ 3968.5 \AA\ \ion{Ca}{ii}\,H} absorption line region of
the summed {\small CORALIE} good spectra for our candidate stars.  The
dotted lines indicate the exact position of the absorption line
centers. A clear reemission feature is visible for {\footnotesize
  HD}\,114386 and {\footnotesize HD}\,147513, whereas only hints of
reemission are observed for {\footnotesize HD}\,65216, {\footnotesize
  HD}\,83443, {\footnotesize HD}\,142415, and {\footnotesize
  HD}\,216770. The spectra have been cleaned as much as possible from
the pollution of the simultaneously recorded thorium-lamp reference
spectra.}
\end{figure*}

\section{Parent star characteristics}

The {\footnotesize CORALIE} planet-search targets have been selected
from the Hipparcos catalogue \citep{ESA-97}. Our planet-star subsample
benefits thus from the photometric and astrometric information
gathered by the satellite. 

A high-resolution spectroscopic abundance study was performed for most
of these stars by N.C.~Santos in his study demonstrating the
metallicity enrichment of stars with planets with regard to comparison
``single'' stars analysed in a homogeneous way \citep[][ and
references therein]{Santos-2001:a,Santos-2003:a}. This study provides
precise values of the effective temperatures, metallicities and
gravity estimates, using a standard local thermodynamical equilibrium
(LTE) analysis. These values have also been updated by using better
recent oscillator strengths ($\log{gf}$) in the procedure
\citep{Santos-2003:c}.  From calibrations of the width and surface of
the {\footnotesize CORALIE} cross-correlation functions \citep[CCF;
described in][]{Santos-2002}, we also derive the stellar projected
rotational velocity $v\sin{i}$ and [Fe/H] when not available from the
spectral analysis.  Like the majority of stars hosting planets, most
of our candidates show a significant metal enrichment compared to the
Sun, with the noticeable exception of {\footnotesize HD}\,6434 (the
second most deficient star hosting a planet known to date) and
{\footnotesize HD}\,111232.

From the colour index, the measured $T_{\rm eff}$ and the
corresponding bolometric correction, we estimate the star
luminosities\footnote{When not available from the spectroscopic study,
  a photometric $T_{\rm eff}$ \citep[following][ not given in
  Table\,\ref{table1}]{Flower-96} is used to derive an indicative star
  luminosity} and we then interpolate the masses and ages in the grid
of Geneva stellar evolutionary models with appropriate metal
abundances \citep{Schaller-92,Schaerer-93}.  For the brightest star in
the sample we estimate, following \citet{Santos-2000:b}, an activity
indicator from the reemission flux in the \ion{Ca}{ii}\,H absorption
line. In some cases, this value is also available in the literature
\citep[mainly in][]{Henry-96}.  Such an indicator is then used to
derive calibrated estimates of the stellar rotational periods and ages
\citep{Noyes-84,Donahue-93}.  Table\,\ref{table1} gathers the
photometric, astrometric, spectroscopic information and inferred
quantities available for our sample of stars hosting exoplanet
candidates.

The activity indicator is, however, not always available in the
literature or cannot be estimated from our spectra because of the star
faintness ({\footnotesize HD}\,65216, {\footnotesize HD}\,114386).
Calcium reemission can then be visually checked on the co-addition of
the {\footnotesize CORALIE} best-S/N spectra.  For our subsample, this
is shown in Fig.\,\ref{fig1} for the \mbox{$\lambda$ 3968.5 \AA\ 
  \ion{Ca}{ii}\,H} absorption line region.  A prominent reemission
feature is clearly visible for {\footnotesize HD}\,114386 and
{\footnotesize HD}\,147513, the latter presenting, moreover, a strong
activity indicator.  However, these stars are only slowly rotating and
no large influence on the radial-velocity measurements is then
expected \citep{Santos-2000:b,Saar-98}.  Hints of reemission are also
observed for {\footnotesize HD}\,65216, {\footnotesize HD}\,83443,
{\footnotesize HD}\,142415, and {\footnotesize HD}\,216770 but again
$v\sin{i}$ is fairly low for these stars for which only a moderate
jitter might be expected.

Some stars in this sample deserve a few more comments:

{\it -- HD\,6434 (HIP\,5054):} The star is classified as subgiant.
Taking into account its low metallicity, the derived absolute
magnitude, effective temperature and gravity estimates also support the
slightly evolved status of the star.

{\it -- HD\,19994 (HIP\,14954, HR\,962, GJ\,128A):} The star is known
to have a close physical M dwarf companion (GJ\,128\,B) a few
arcseconds away \citep[$\sim$\,100\,AU;][]{Hale-94}. Planets in
binary systems are important for our understanding of planet formation
because they seem to present different mass and orbital properties
than planets orbiting single stars
\citep{Zucker-2002,Eggenberger-2003,Udry-2003:a}.  The rotational
velocity of {\footnotesize HD}\,19994 is fairly large
($v\sin{i}$\,=\,8.1\,km\,s$^{-1}$) thus, even if the star is not
clearly active, some small radial-velocity jitter might be expected.
Contamination from the secondary could also be a concern although the
6.4 magnitude difference between the stars makes it very unlikely.

{\it -- HD\,82943 (HIP\,47007):} A fair amount of the easy-to-burn
element $^6$Li has been detected in the spectra of this star
\citep{Israelian-2001,Israelian-2003} suggesting a planet engulfment
in the late stages of the system formation, after the star has reached
the Main Sequence. The not too deep convective zone of this G0 dwarf
allows then for the survival of this element.

{\it -- HD\,111232 (HIP\,62534):} The high velocity
($V$\,=\,104.4\,km\,s$^{-1}$) and low metallicity ([Fe/H]\,=\,$-0.36$)
of the star indicate that it probably belongs to the thick disk
population.  {\footnotesize HD}\,111232 is a proposed binary in the
Hipparcos catalogue \citep{ESA-97}.  It is one of the so-called {\sl
  problem stars} for which an astrometric acceleration solution is
provided (flagged ``G'' in the H59 field of the main catalogue).
However, no companion with $\Delta V$\,$\leq$\,3.0 was found close to
the star (within 1.08\,${\arcsec}$) by speckle interferometry
\citep{Mason-98}.

{\it -- HD\,121504 (HIP\,68162):} A visual companion ({\footnotesize
  CPD}\,$-55$:5793) is observed at a separation of 34.2\,${\arcsec}$
\citep{Dommanget-94}.  It is an A2 star of magnitude $V$\,=\,9.17 with
different proper motions than {\footnotesize HD}\,121504. The pair is
thus not physical. No companion was found close to the star by speckle
interferometry \citep{Mason-98}.

{\it -- HD\,142415 (HIP\,78169):} A Rosat-All-Sky-Survey X-ray source
(1RXS\,J155740.7-601154) is observed close to the star, at
$\sim$\,5\,${\arcsec}$. No bright companion was found close to the star
by speckle interferometry measurements \citep{Mason-98}.

{\it -- HD\,147513 (HIP\,80337, HR\,6094, GJ\,620.1A):} The star was
proposed to be a barium dwarf by \citet{Porto-97}. A common proper
motion white dwarf (WD; $V$\,=\,10), at a projected distance of
5360\,AU, could support the explanation of the origin of the barium
feature by the process of mass transfer in a binary system, in which
the secondary component accreted matter from the evolved primary, now
the WD \citep[see e.g.][ and references therein]{Jorissen-98}.  To
account for the observed large separation, \citet{Porto-97} invoke a
possible ejection of the WD from an originally quadruple system also
including {\footnotesize HD}\,147513. In such a case, this would be
the first known case of a planet orbiting a solar-type star in a
binary system with the stellar companion in the last stage of its
life. The influence on the planet evolution of the mass transfer
between two stars is still poorly studied.  The mentioned authors also
include the star in the 0.3\,Gyr old Ursa Major kinematical group.
Although some doubts (but no rejection) were cast on the membership
\citep{King-2003}, it is worth noticing that the activity level,
activity-derived age and metallicity (Table\,\ref{table1}) correspond
well to the Ursa Major group.  Finally, no bright companion was found
closer-in to the star by speckle interferometry measurements
\citep{Mason-98}.

\section{Radial-velocity data and orbital solutions} 

The radial-velocity measurements presented in this paper were obtained
with the {\footnotesize CORALIE} echelle spectrograph mounted on the
1.2-m Euler Swiss telescope at La~Silla Observatory (ESO, Chile).
{\footnotesize CORALIE} is a similar but improved version of the
{\footnotesize ELODIE} spectrograph at the Haute-Provence Observatory
(CNRS, France).  Details on the instrument design, reduction
procedures as well as radial-velocity estimates based on simultaneous
thorium-lamp measurements and cross-correlation technique can be found
in \citet{Baranne-96}.

\begin{table*}[t!]
\caption{
\label{table2}
{\footnotesize CORALIE} best Keplerian orbital solutions
as well as inferred planetary parameters for the 1-planet systems.
$Span$ is the time interval in days between the first and last
measurements. $\sigma(O-C)$ is the weighted r.m.s. of the residuals
around the derived solutions. $mask$ is the template used in the
cross-correlation scheme for the radial-velocity estimate
(see Sect.\,3). $Det.Ref.$ is the reference to the planet detection 
announcement (conference or press release).}
\begin{tabular}{l@{}lr@{\,$\pm$\,}lr@{\,$\pm$\,}lr@{\,$\pm$\,}lr@{\,$\pm$\,}lr@{\,$\pm$\,}l}
\hline
  \multicolumn{2}{l}{\bf Parameter} 
& \multicolumn{2}{c}{\bf {\footnotesize HD}\,6434\,b} 
& \multicolumn{2}{c}{\bf {\footnotesize HD}\,19994\,b} 
& \multicolumn{2}{c}{\bf {\footnotesize HD}\,65216\,b} 
& \multicolumn{2}{c}{\bf {\footnotesize HD}\,83443\,b} 
& \multicolumn{2}{c}{\bf {\footnotesize HD}\,92788\,b} \\
& &\multicolumn{2}{c}{ {\footnotesize HIP}\,5054\,b} 
& \multicolumn{2}{c}{ {\footnotesize HIP}\,14954\,b} 
& \multicolumn{2}{c}{ {\footnotesize HIP}\,38558\,b} 
& \multicolumn{2}{c}{ {\footnotesize HIP}\,47202\,b} 
& \multicolumn{2}{c}{ {\footnotesize HIP}\,52409\,b} \\
\hline
$Det.Ref.$ &    &\multicolumn{2}{c}{Queloz et al.}
                &\multicolumn{2}{c}{Queloz et al.}  
                &\multicolumn{2}{c}{Udry et al.}  
                &\multicolumn{2}{c}{Mayor et al.}  
                &\multicolumn{2}{c}{Queloz et al.$^{\dagger}$}  \\
           &    &\multicolumn{2}{c}{(2000)}  
                &\multicolumn{2}{c}{(2000)}  
                &\multicolumn{2}{c}{(2003b)}  
                &\multicolumn{2}{c}{(2000)}  
                &\multicolumn{2}{c}{(2000)}  \\
\hline
$P$ &$[$days$]$     &21.998   &0.009  &535.7   &3.1   &613.1  &11.4  
                    &2.98565  &0.00003 &325.0  &0.5  \\
$T$ &[JD-2\,400\,000]       &51490.8  &0.6    
                    &50944     &12    &50762  &25 
                    &51497.5   &0.3   &51090.3 &3.5   \\ 
$e$ &               &0.17      &0.03  &0.30   &0.04  &0.41   &0.06
                    &0.013     &0.013 &0.35   &0.01   \\   
$V$ &[km\,s$^{-1}$] &23.023   &0.001  &19.335 &0.001 &42.674 &0.002
                    &29.027   &0.001  &$-4.467$ &0.001 \\
$\omega$ &[deg]     &156      &11     &41     &8     &198    &6
                    &11       &11     &279    &3  \\
$K$ &[m\,s$^{-1}$]  &34.2     &1.1    &36.2   &1.9   &33.7   &1.1 
                    &58.1     &0.4    &106.2  &1.8   \\
$a_1\sin i$ &[$\mathrm{10^{-3}}$\,AU]   
                &\multicolumn{2}{c}{0.068}  &\multicolumn{2}{c}{1.701} 
                &\multicolumn{2}{c}{1.729}  &\multicolumn{2}{c}{1.594} 
                &\multicolumn{2}{c}{2.966}   \\
$f(m)$ &$\mathrm{[10^{-9}\,M_{\odot}]}$ 
                &\multicolumn{2}{c}{0.087}  &\multicolumn{2}{c}{2.289}   
                &\multicolumn{2}{c}{1.835}  &\multicolumn{2}{c}{6.062}
                &\multicolumn{2}{c}{32.94}  \\
$m_{2}\,\sin i$ &$\mathrm{[M_{\rm Jup}]}$ 
                &\multicolumn{2}{c}{0.39} &\multicolumn{2}{c}{1.68}
                &\multicolumn{2}{c}{1.21} &\multicolumn{2}{c}{0.38}
                &\multicolumn{2}{c}{3.58} \\
$a$ &[AU]       &\multicolumn{2}{c}{0.14} &\multicolumn{2}{c}{1.42} 
                &\multicolumn{2}{c}{1.37} &\multicolumn{2}{c}{0.039} 
                &\multicolumn{2}{c}{0.96} \\
\hline
$N_{\rm meas}$ &    &\multicolumn{2}{c}{130}  &\multicolumn{2}{c}{48} 
                    &\multicolumn{2}{c}{70}   &\multicolumn{2}{c}{257}
                    &\multicolumn{2}{c}{55}   \\
$Span$ &[days]      &\multicolumn{2}{c}{1501} &\multicolumn{2}{c}{1519} 
                    &\multicolumn{2}{c}{1460} &\multicolumn{2}{c}{1455}
                    &\multicolumn{2}{c}{1451} \\
$\sigma (O-C)$\,\,  &[m\,s$^{-1}$] 
                    &\multicolumn{2}{c}{10.6} &\multicolumn{2}{c}{8.1} 
                    &\multicolumn{2}{c}{6.8}  &\multicolumn{2}{c}{9.0} 
                    &\multicolumn{2}{c}{8.0}  \\
$mask$ & &\multicolumn{2}{c}{weighted $K0$} 
         &\multicolumn{2}{c}{$K0$}
         &\multicolumn{2}{c}{weighted $K0$} 
         &\multicolumn{2}{c}{weighted $K0$} 
         &\multicolumn{2}{c}{weighted $K0$}\\
\hline\\
\hline
  \multicolumn{2}{l}{\bf Parameter} 
& \multicolumn{2}{c}{\bf {\footnotesize HD}\,111232\,b}
& \multicolumn{2}{c}{\bf {\footnotesize HD}\,114386\,b}
& \multicolumn{2}{c}{\bf {\footnotesize HD}\,121504\,b}
& \multicolumn{2}{c}{\bf {\footnotesize HD}\,142415\,b} 
& \multicolumn{2}{c}{\bf {\footnotesize HD}\,147513\,b} \\
& & \multicolumn{2}{c}{ {\footnotesize HIP}\,62534\,b}
& \multicolumn{2}{c}{ {\footnotesize HIP}\,64295\,b}
& \multicolumn{2}{c}{ {\footnotesize HIP}\,68162\,b}
& \multicolumn{2}{c}{ {\footnotesize HIP}\,78169\,b} 
& \multicolumn{2}{c}{ {\footnotesize HIP}\,80337\,b} \\
\hline
$Det.Ref.$ &    &\multicolumn{2}{c}{Udry et al.}
                &\multicolumn{2}{c}{Udry et al.}  
                &\multicolumn{2}{c}{Queloz et al.}  
                &\multicolumn{2}{c}{Udry et al.}  
                &\multicolumn{2}{c}{Udry et al.}  \\
           &    &\multicolumn{2}{c}{(2003b)}  
                &\multicolumn{2}{c}{(2002)}  
                &\multicolumn{2}{c}{(2000)}  
                &\multicolumn{2}{c}{(2003b)}  
                &\multicolumn{2}{c}{(2002)}  \\
\hline
$P$ &$[$days$]$     &1143    &14    &937.7   &15.6  
                    &63.33   &0.03  &386.3   &1.6   &528.4  &6.3 \\ 
$T$ &[JD-2\,400\,000]     &51230   &20    &50454   &43 
                    &51450   &2     &51519   &4     &51123  &20  \\ 
$e$ &               &0.20   &0.01   &0.23    &0.03
                    &0.03   &0.01   &\multicolumn{2}{c}{0.5 (fixed)} 
                    &0.26   &0.05  \\   
$V$ &[km\,s$^{-1}$] &104.4  &0.001  &33.370  &0.001
                    &19.617 &0.001  &$-11.811$ &0.001 &12.924 &0.001 \\
$\omega$ &[deg]     &98     &6      &273     &14
                    &265    &12     &255     &4     &282  &9 \\
$K$ &[m\,s$^{-1}$]  &159.3  &2.3    &34.3    &1.6
                    &55.8   &0.9    &51.3    &2.3   &29.3 &1.8   \\
$a_1\sin i$ &[$\mathrm{10^{-3}}$\,AU]   
                &\multicolumn{2}{c}{16.39}  &\multicolumn{2}{c}{2.875}
                &\multicolumn{2}{c}{0.325}  &\multicolumn{2}{c}{1.579}
                &\multicolumn{2}{c}{1.377}  \\
$f(m)$ &$\mathrm{[10^{-9}\,M_{\odot}]}$ 
                &\multicolumn{2}{c}{450.0}  &\multicolumn{2}{c}{3.604}
                &\multicolumn{2}{c}{1.140}  &\multicolumn{2}{c}{3.517}
                &\multicolumn{2}{c}{1.248}  \\
$m_{2}\,\sin i$ &$\mathrm{[M_{\rm Jup}]}$ 
                &\multicolumn{2}{c}{6.80}   &\multicolumn{2}{c}{1.24}
                &\multicolumn{2}{c}{1.22}   &\multicolumn{2}{c}{1.62} 
                &\multicolumn{2}{c}{1.21}   \\
$a$ &[AU]       &\multicolumn{2}{c}{1.97}   &\multicolumn{2}{c}{1.65} 
                &\multicolumn{2}{c}{0.33}   &\multicolumn{2}{c}{1.05} 
                &\multicolumn{2}{c}{1.32}   \\
\hline
$N_{\rm meas}$ &    &\multicolumn{2}{c}{38}  &\multicolumn{2}{c}{58} 
                    &\multicolumn{2}{c}{100} &\multicolumn{2}{c}{137} 
                    &\multicolumn{2}{c}{30} \\
$Span$ &[days]      &\multicolumn{2}{c}{1181} &\multicolumn{2}{c}{1550} 
                    &\multicolumn{2}{c}{1496} &\multicolumn{2}{c}{1529}
                    &\multicolumn{2}{c}{1690} \\
$\sigma (O-C)$\,\,  &[m\,s$^{-1}$] 
                    &\multicolumn{2}{c}{7.5} &\multicolumn{2}{c}{10.2} 
                    &\multicolumn{2}{c}{11.6} &\multicolumn{2}{c}{10.6} 
                    &\multicolumn{2}{c}{5.7} \\ 
$mask$  &           &\multicolumn{2}{c}{weighted $K0$}
                    &\multicolumn{2}{c}{weighted $K0$}
                    &\multicolumn{2}{c}{weighted $K0$}
                    &\multicolumn{2}{c}{$K0$} 
                    &\multicolumn{2}{c}{weighted $K0$}\\
\hline \\
\end{tabular}
\begin{tabular}{l@{}lr@{\,$\pm$\,}lr@{\,$\pm$\,}lr@{\,$\pm$\,}l}
\hline
  \multicolumn{2}{l}{\bf Parameter} 
& \multicolumn{2}{c}{\bf {\footnotesize HD}\,196050\,b}
& \multicolumn{2}{c}{\bf {\footnotesize HD}\,216437\,b}
& \multicolumn{2}{c}{\bf {\footnotesize HD}\,216770\,b} \\
& & \multicolumn{2}{c}{ {\footnotesize HIP}\,101806\,b}
& \multicolumn{2}{c}{ {\footnotesize HIP}\,113137\,b}
& \multicolumn{2}{c}{ {\footnotesize HIP}\,113238\,b} \\
\hline
$Det.Ref.$ &      &\multicolumn{2}{c}{\citet{Udry-2002:b}$^{\ddagger}$}  
                &\multicolumn{2}{c}{\citet{Udry-2002:b}$^{\ddagger}$}  
                &\multicolumn{2}{c}{\citet{Udry-2003:e}}  \\
\hline
$P$ &$[$days$]$     &1321     &54  
                    &1256     &35    &118.45 &0.55  \\ 
$T$ &[JD-2\,400\,000]      &52045  &66 
                    &50693    &130   &52672  &3.5  \\ 
$e$ &               &\multicolumn{2}{c}{0.3 (fixed)}
                    &0.29     &0.12  &0.37   &0.06 \\   
$V$ &[km\,s$^{-1}$] &61.342   &0.005
                    &$-2.278$ &0.004 &31.153 &0.002 \\
$\omega$ &[deg]     &147      &12   
                    &63       &22    &281    &10 \\
$K$ &[m\,s$^{-1}$]  &55.0     &6.2 
                    &34.6     &5.7   &30.9   &1.9  \\
$a_1\sin i$ &[$\mathrm{10^{-3}}$\,AU]   
                &\multicolumn{2}{c}{6.367}
                &\multicolumn{2}{c}{3.815}   &\multicolumn{2}{c}{0.313}\\
$f(m)$ &$\mathrm{[10^{-9}\,M_{\odot}]}$ 
                &\multicolumn{2}{c}{19.74}
                &\multicolumn{2}{c}{4.694}   &\multicolumn{2}{c}{0.291}\\
$m_{2}\,\sin i$ &$\mathrm{[M_{\rm Jup}]}$ 
                &\multicolumn{2}{c}{3.02}
                &\multicolumn{2}{c}{1.82}    &\multicolumn{2}{c}{0.65} \\
$a$ &[AU]       &\multicolumn{2}{c}{2.43} 
                &\multicolumn{2}{c}{2.32}    &\multicolumn{2}{c}{0.46} \\
\hline
$N_{\rm meas}$ &    &\multicolumn{2}{c}{31} 
                    &\multicolumn{2}{c}{21}   &\multicolumn{2}{c}{16} \\
$Span$ &[days]      &\multicolumn{2}{c}{1364} 
                    &\multicolumn{2}{c}{1405} &\multicolumn{2}{c}{827} \\
$\sigma (O-C)$\,\,  &[m\,s$^{-1}$] 
                    &\multicolumn{2}{c}{7.2} 
                    &\multicolumn{2}{c}{7.2} &\multicolumn{2}{c}{7.8} \\ 
$mask$   &          &\multicolumn{2}{c}{weighted $K0$}
                    &\multicolumn{2}{c}{weighted $K0$} 
                    &\multicolumn{2}{c}{weighted $K0$} \\
\hline
\end{tabular}

$^{\dagger}$ Independently discovered by \citet{Fischer-2001}
\hspace{0.5cm}
$^{\ddagger}$ Independently discovered by \citet{Jones-2002} 
\end{table*}

The typical precision obtained with {\footnotesize CORALIE} is
$\sim$\,3\,m\,s$^{-1}$ for bright stars \citep{Queloz-2001:b}.
However, due to the small size of the telescope, our sample stars
\citep{Udry-2000:a} are mainly photon-noise limited. For the
derivation of the orbital solution an {\sl instrumental} error of
3\,m\,s$^{-1}$ is quadratically added to photon noise. Then we only
take into account the spectra with good signal-to-noise, typically
corresponding to radial-velocity uncertainties below 10 to
15\,m\,s$^{-1}$, depending on the star.  Also, to improve the Doppler
information extraction from the spectra, a new weighted
cross-correlation scheme was developed \citep{Pepe-2002:a}. It is
applied for the velocity estimates of most of the stars of this
paper\footnote{The CORALIE numerical mask used in the cross
  correlation for the velocity estimate was built from a K0-dwarf
  spectrum. For much earlier spectral-type stars (typically
  $\leq$\,G1), the mismatch between the stellar spectrum and the
  template is enhanced by the weighted scheme and no improvement is
  obtained for the radial-velocity measurements.  The usual cross
  correlation scheme is then used in such cases (Table\,\ref{table2})}.

\begin{figure*}[th!]
\begin{center}
\psfig{width=0.418\hsize,file=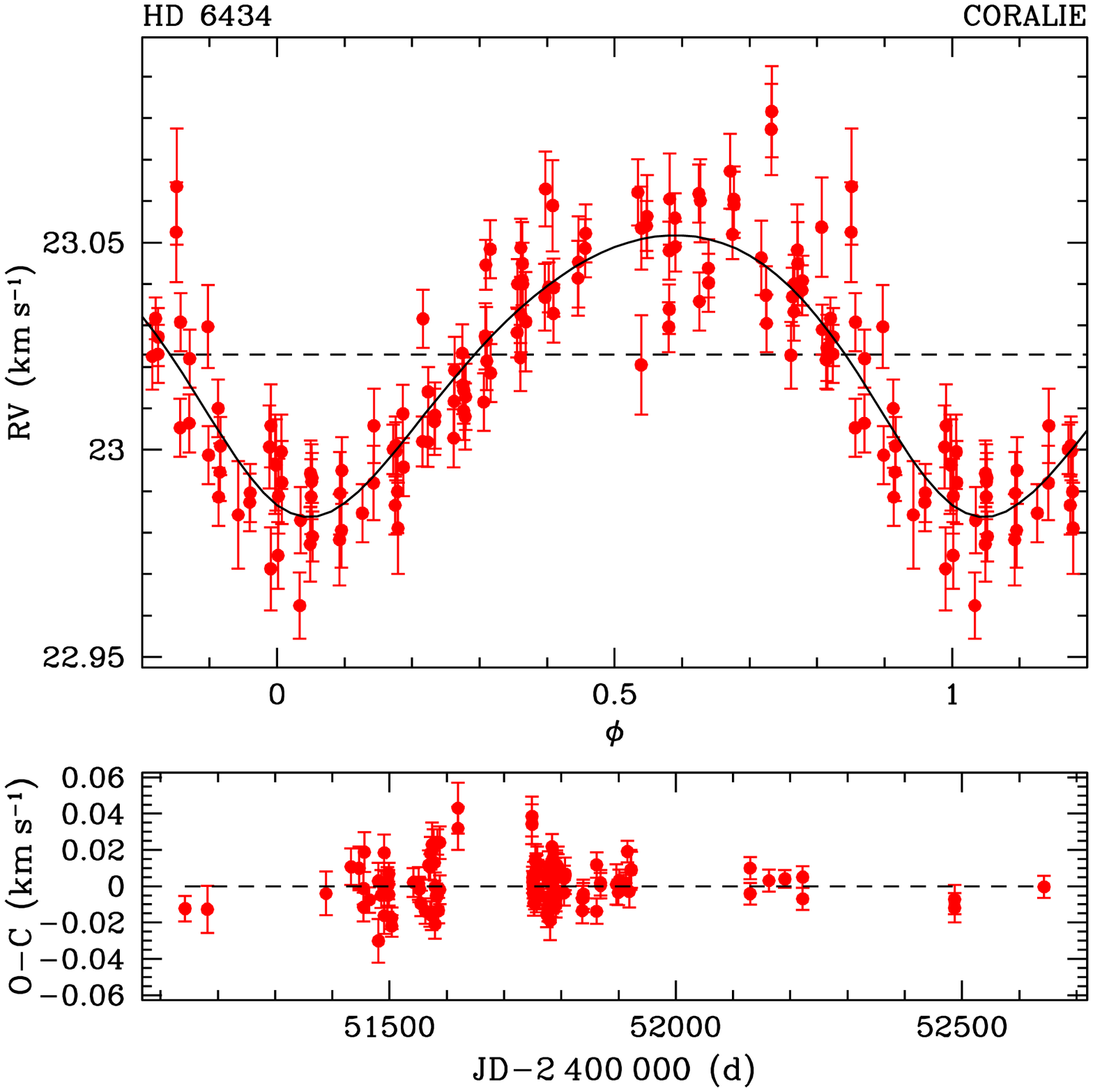}
\psfig{width=0.418\hsize,file=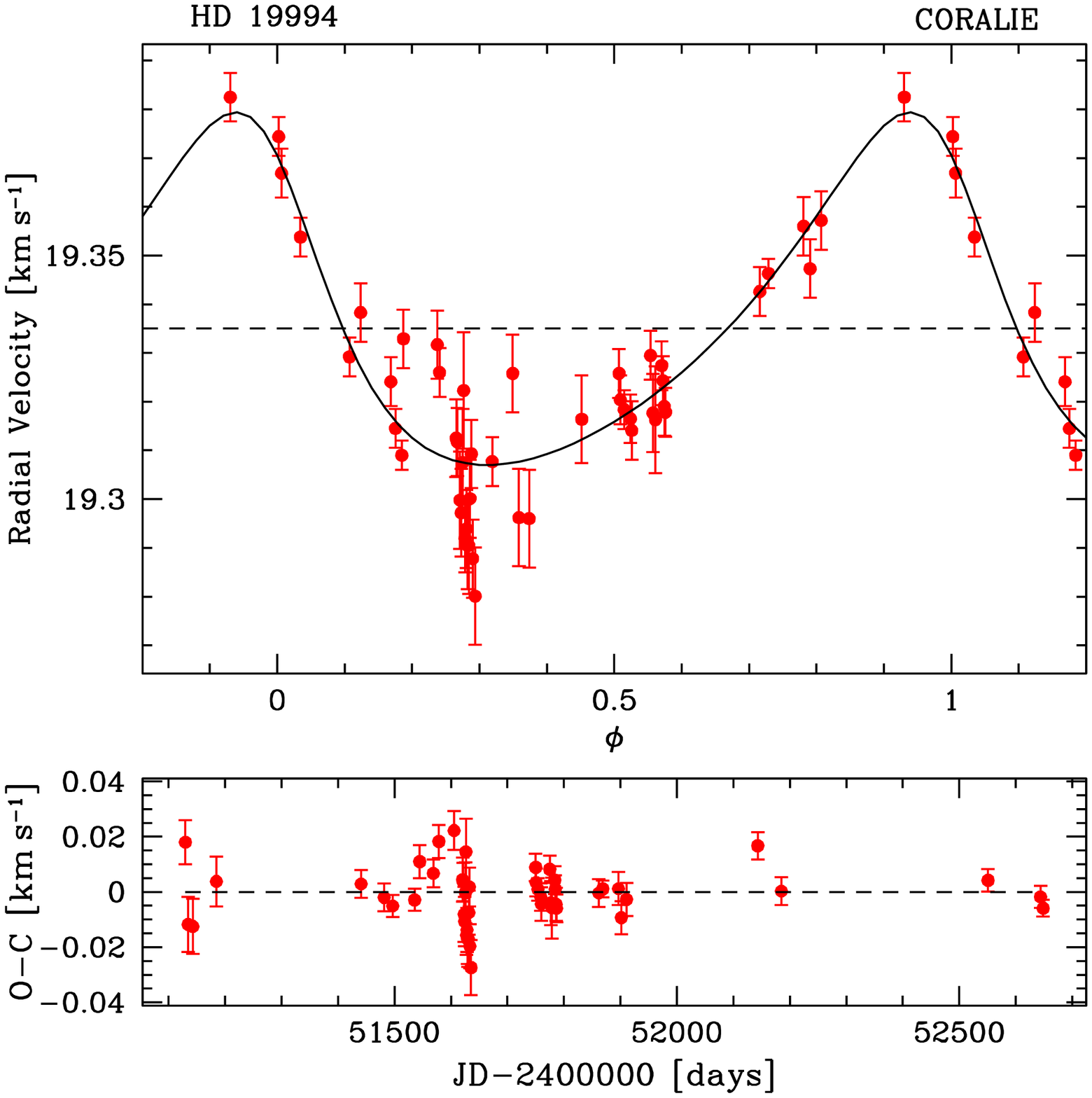}

\psfig{width=0.418\hsize,file=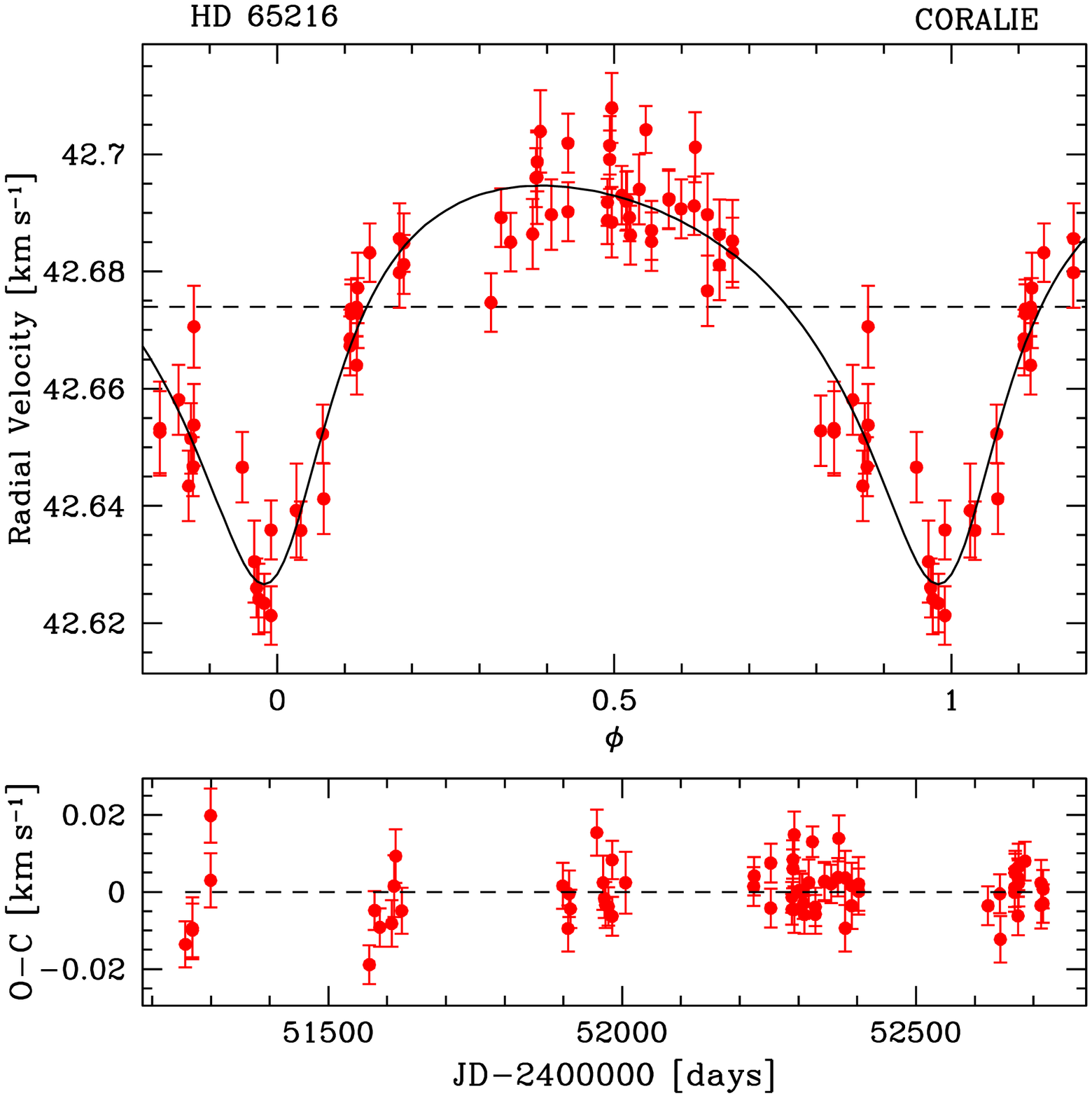}
\psfig{width=0.418\hsize,file=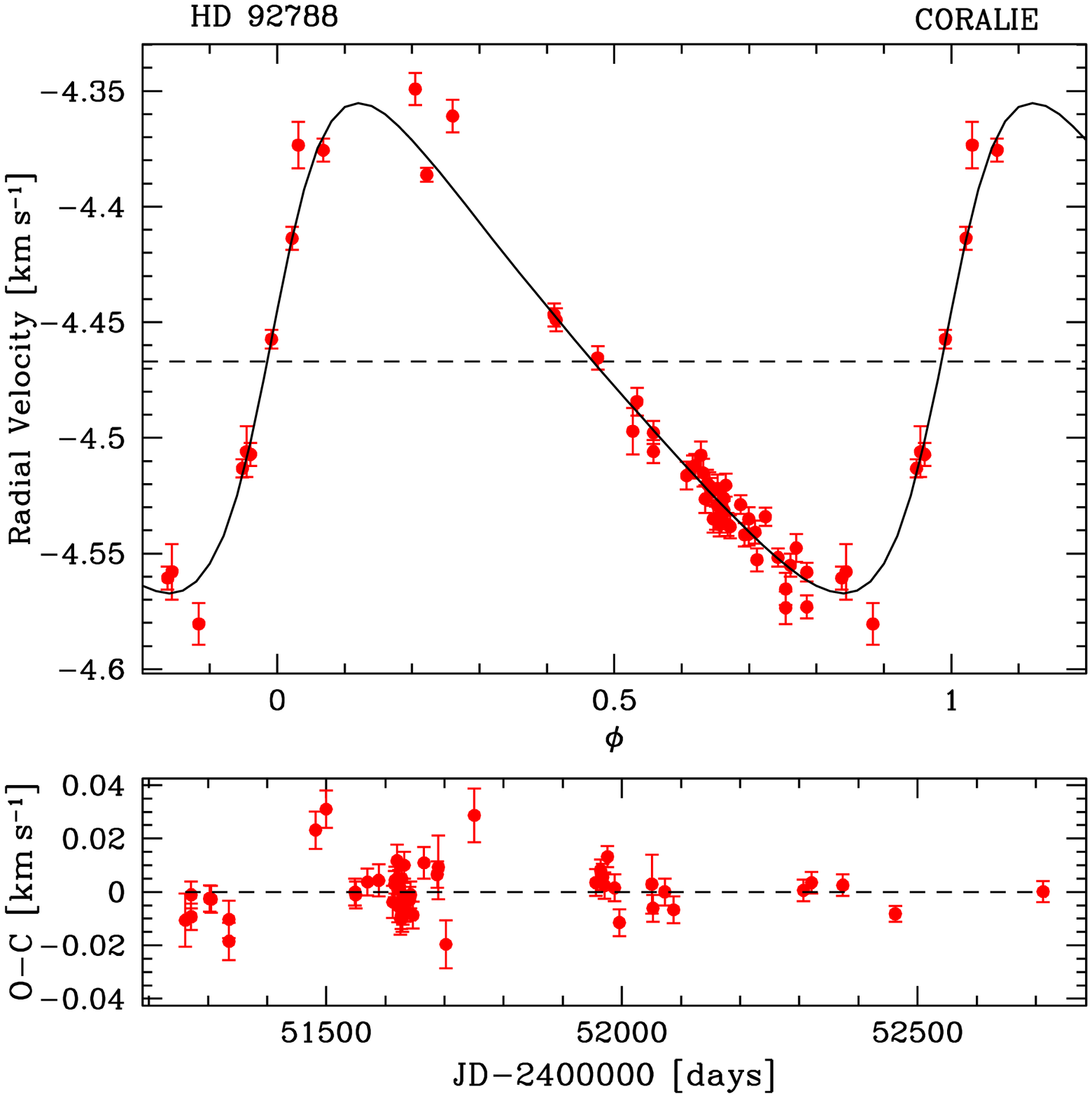}

\psfig{width=0.418\hsize,file=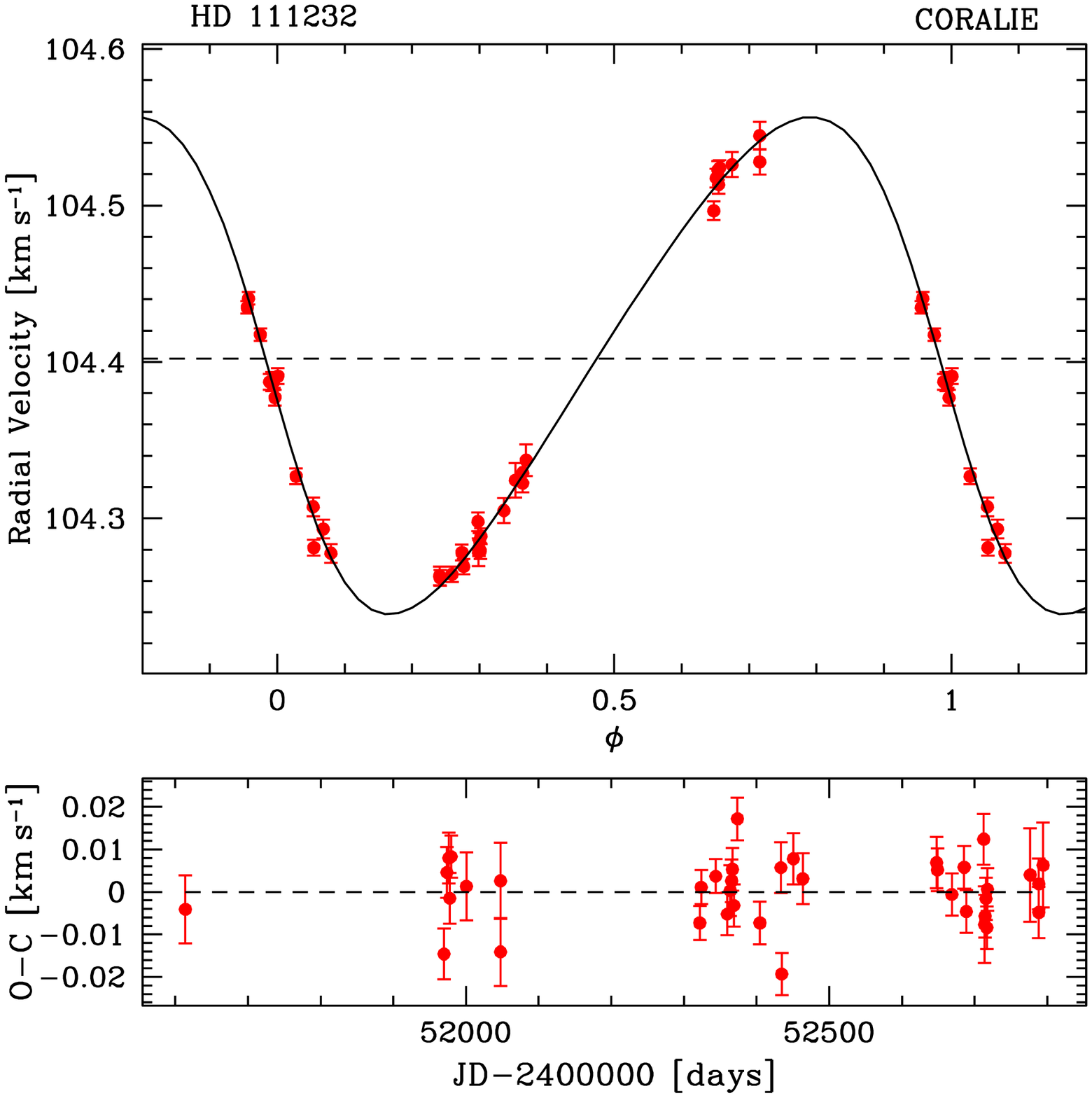}
\psfig{width=0.418\hsize,file=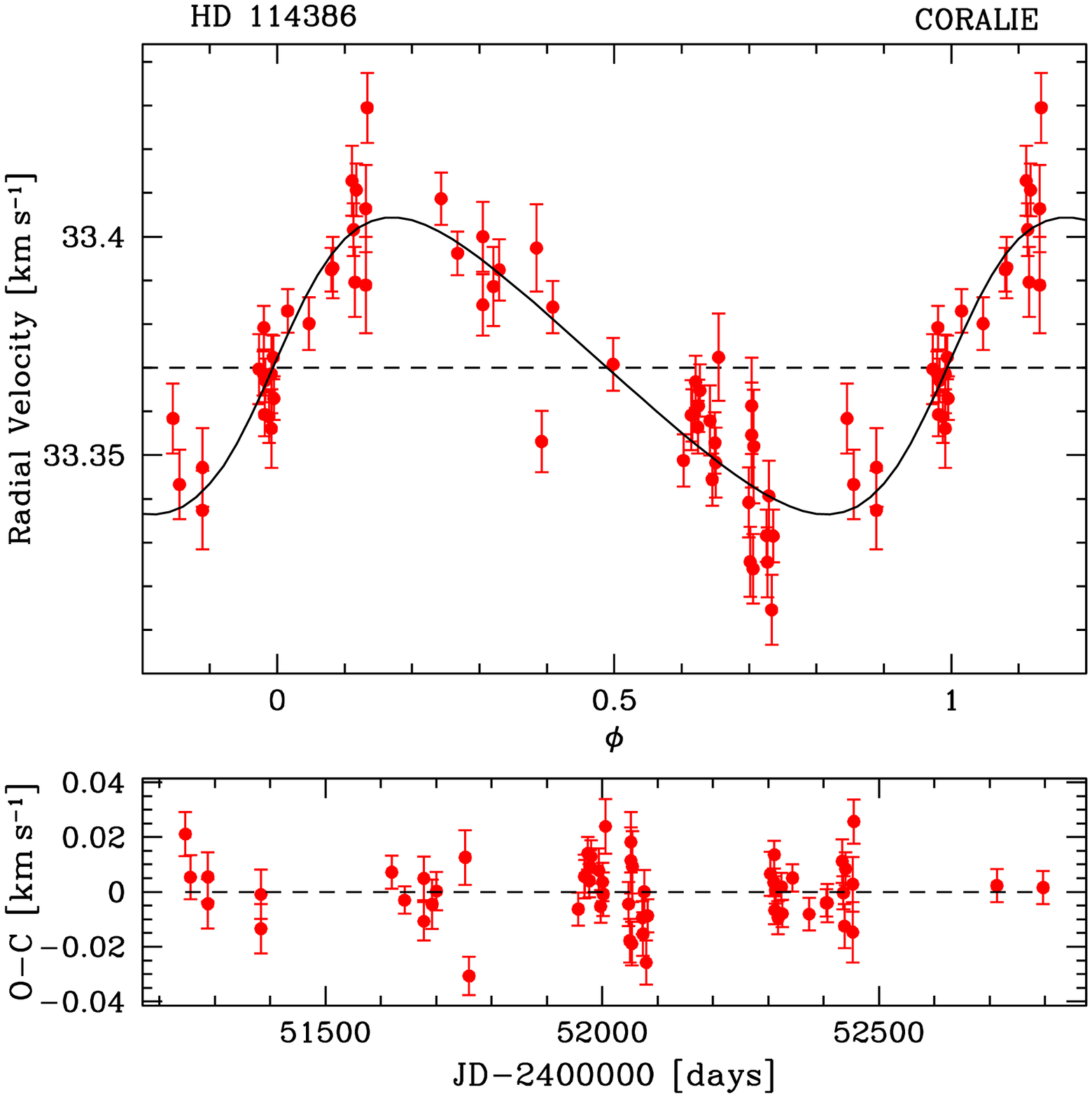}
\caption{
\label{fig2}
Phase-folded radial-velocity measurements obtained with {\footnotesize
  CORALIE} for {\footnotesize HD}\,6434, {\footnotesize HD}\,19994,
{\footnotesize HD}\,65216, {\footnotesize HD}\,92788, {\footnotesize
  HD}\,111232 and {\footnotesize HD}\,114386, superimposed on the best
Keplerian planetary solution (top panel in each diagram). The
residuals as a function of Julian Date are displayed in the lower
panels.}
\end{center}
\end{figure*}

\begin{figure*}[th!]
\begin{center}
\psfig{width=0.418\hsize,file=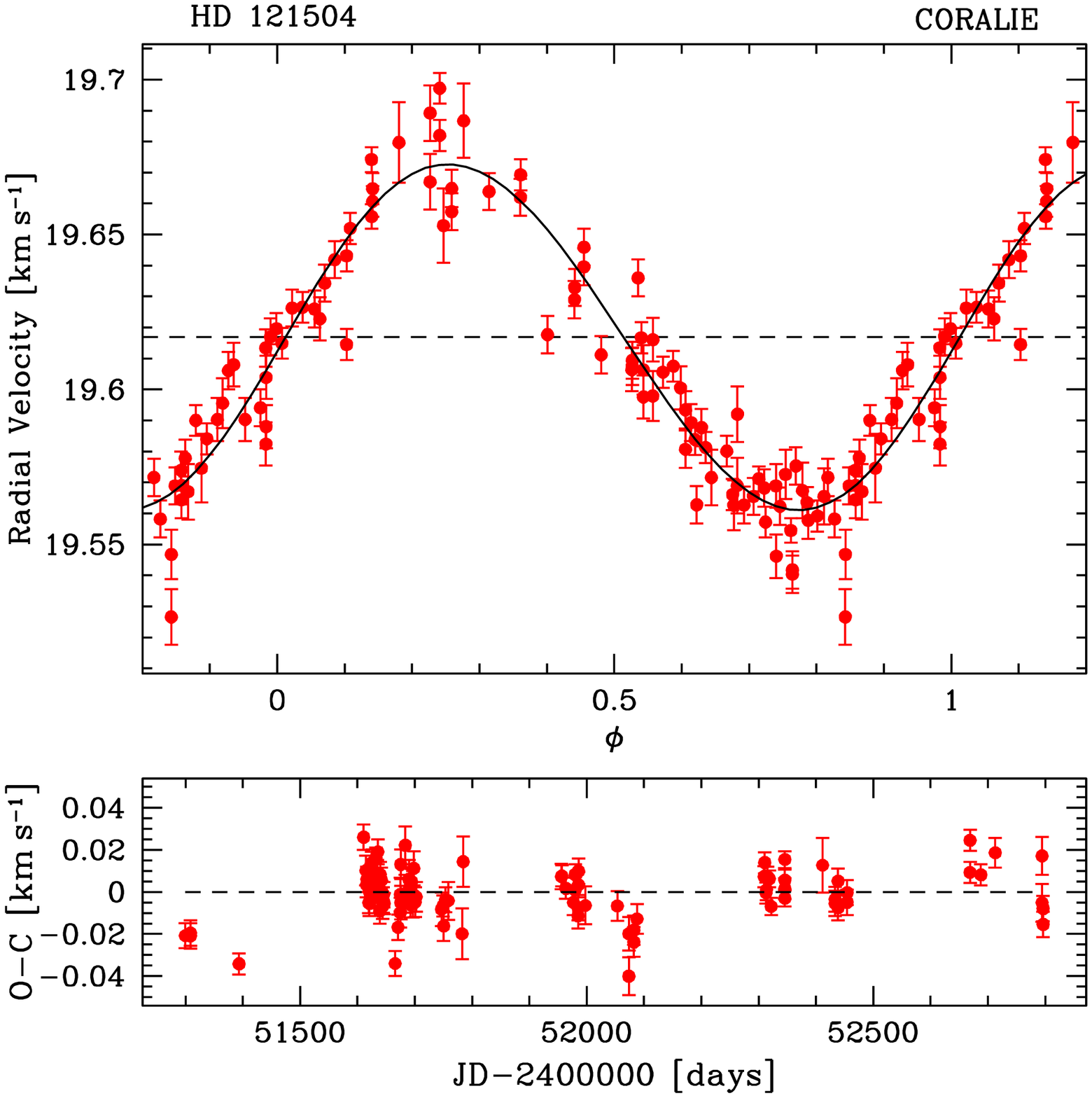}
\psfig{width=0.418\hsize,file=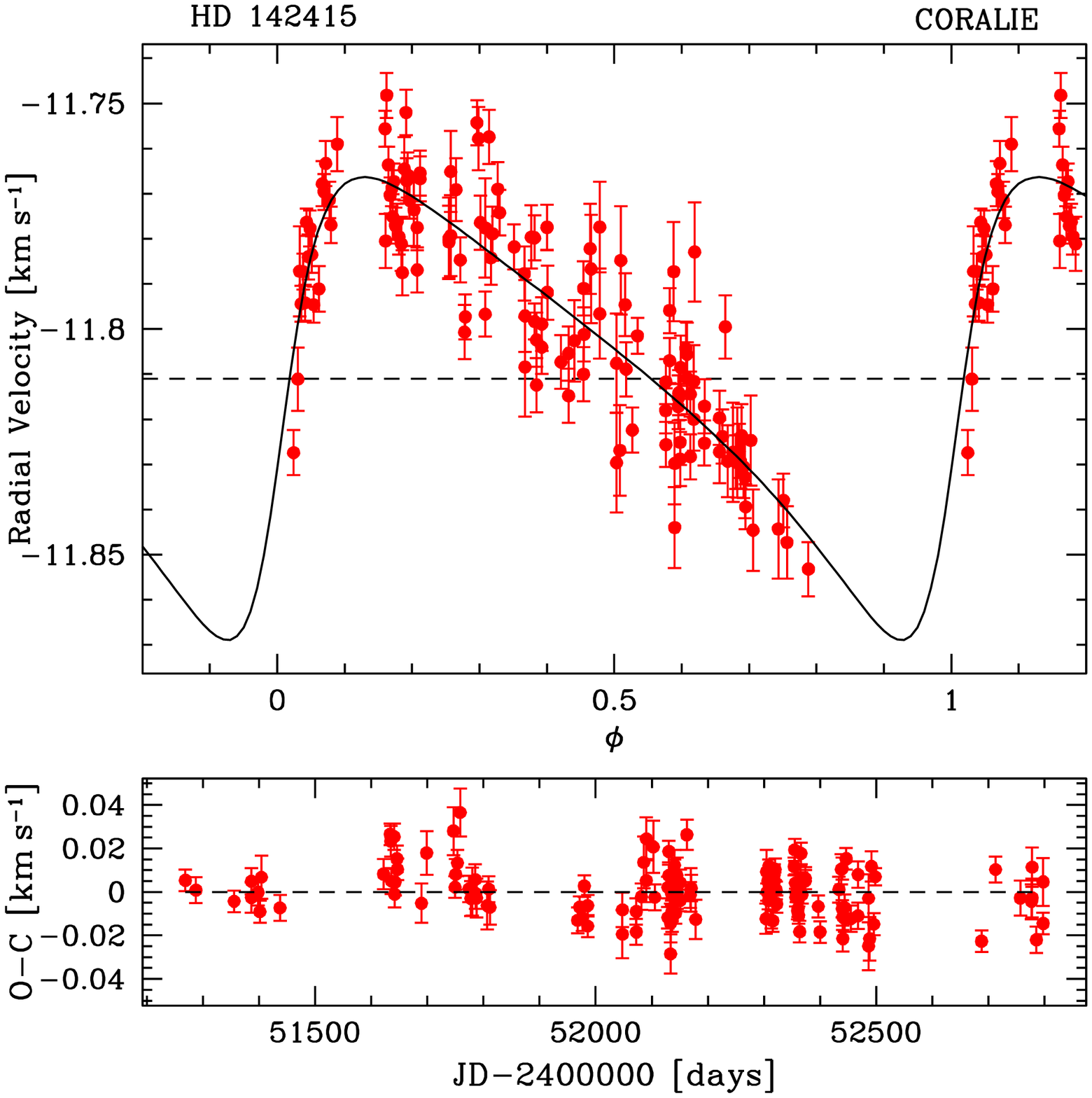}

\psfig{width=0.418\hsize,file=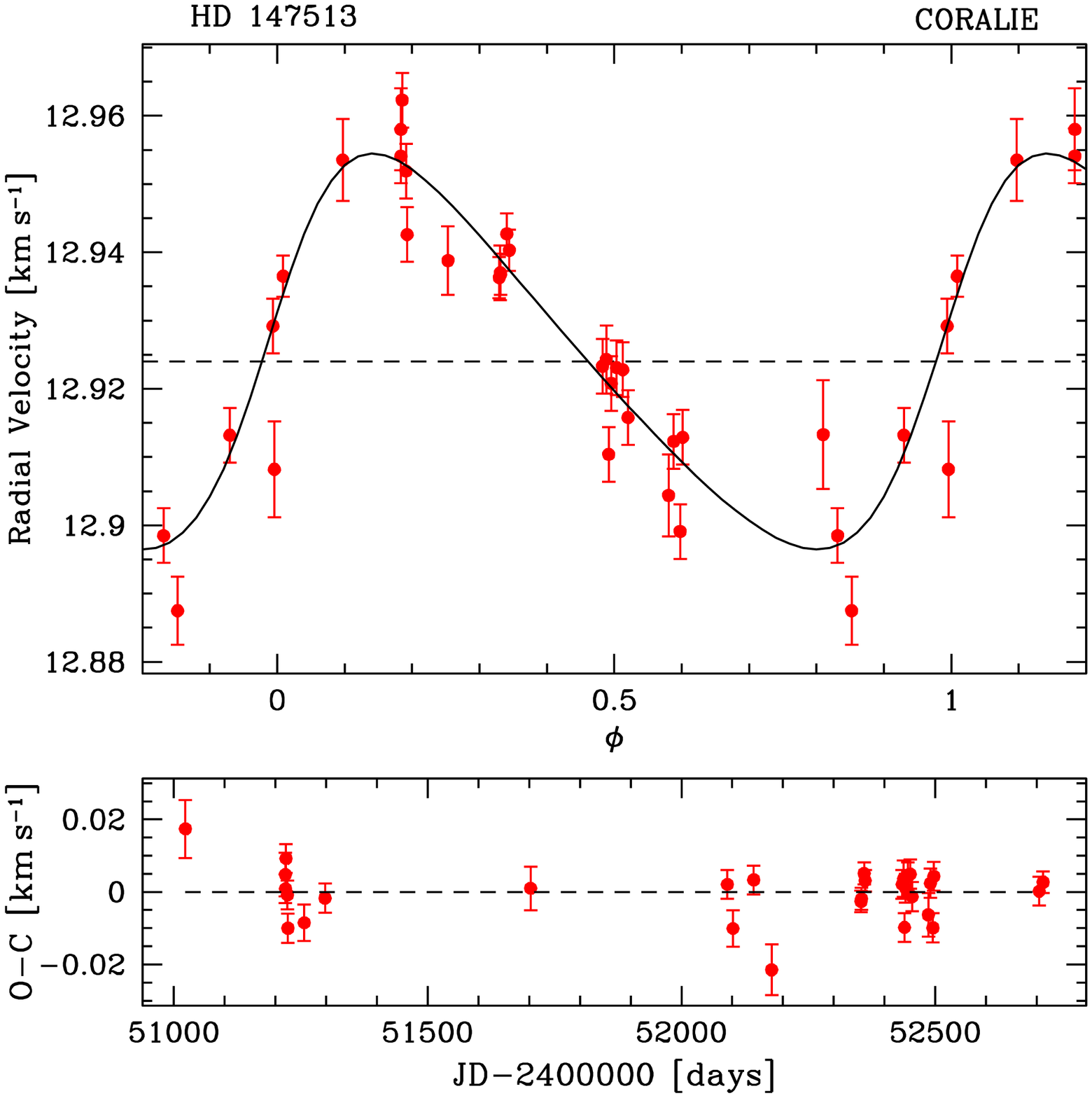}
\psfig{width=0.418\hsize,file=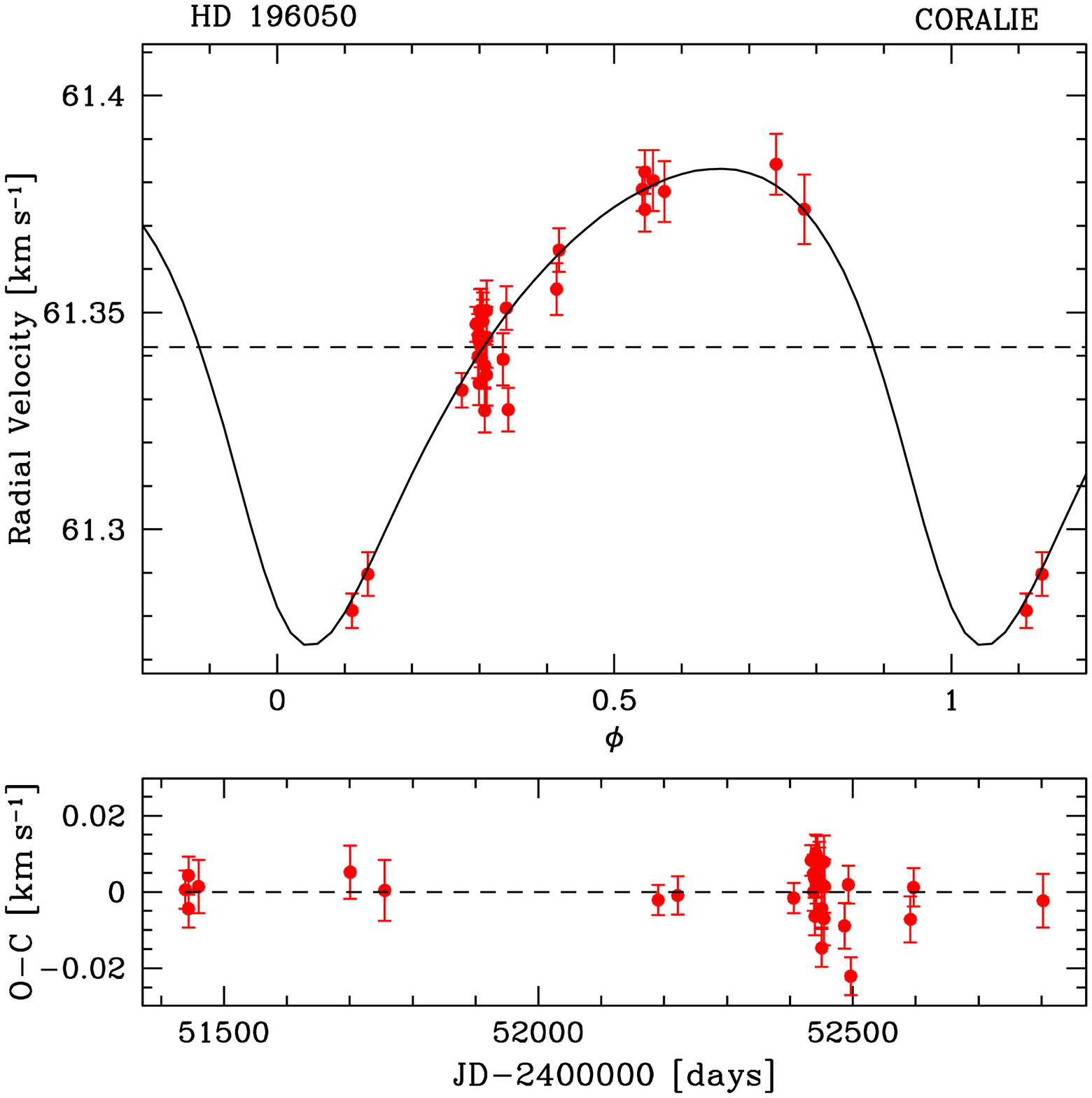}

\psfig{width=0.418\hsize,file=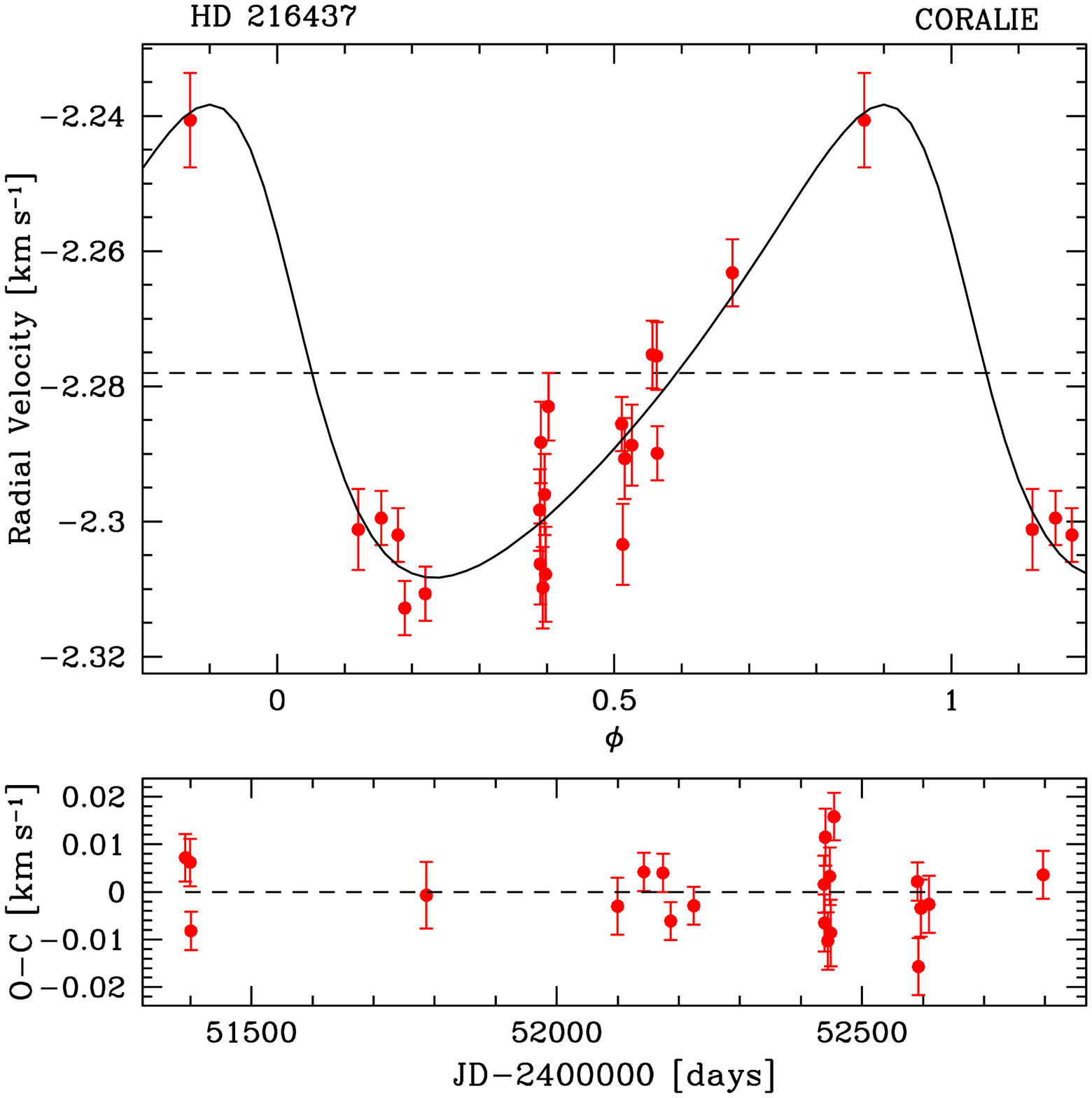}
\psfig{width=0.418\hsize,file=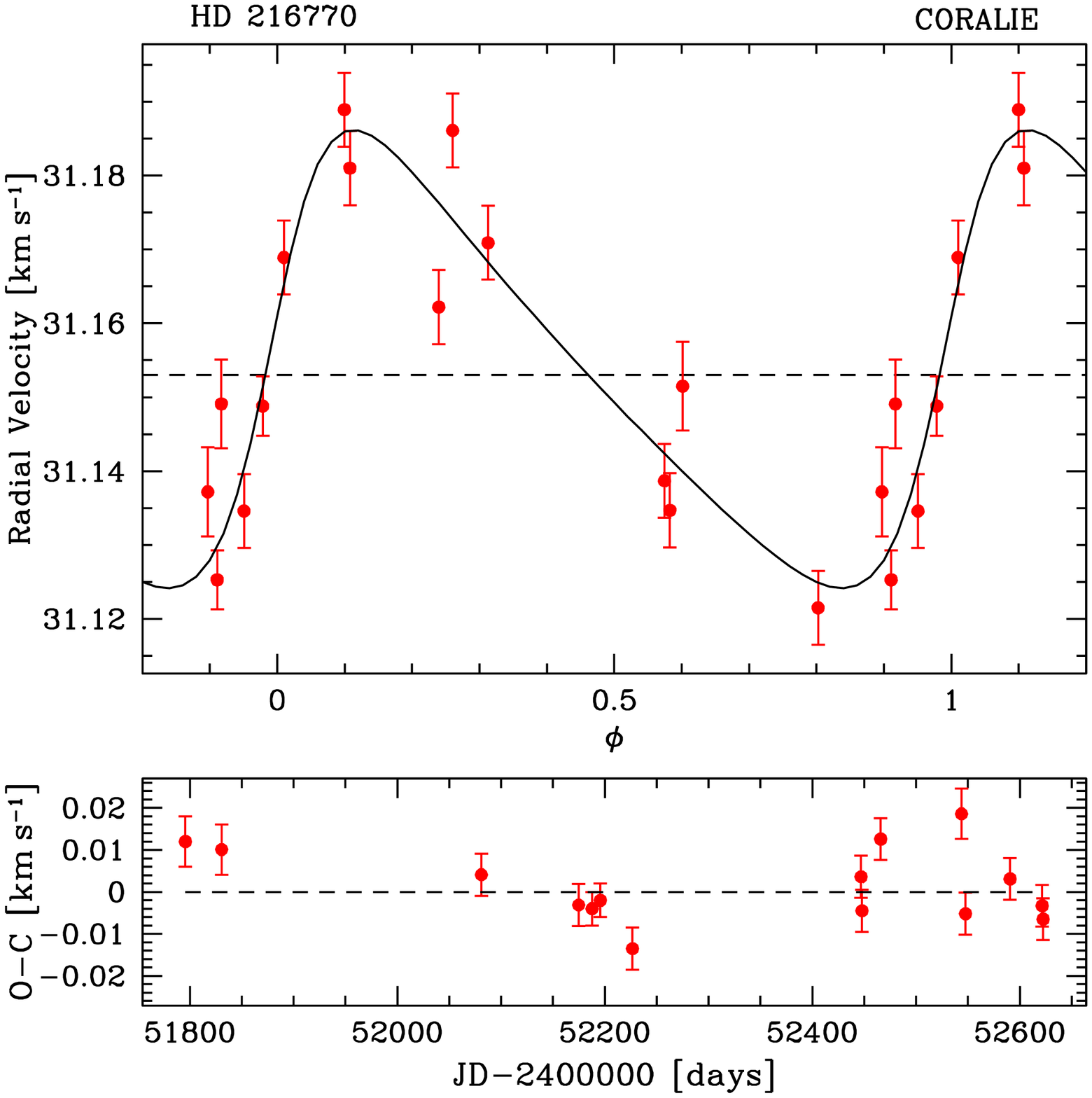}
\caption{
\label{fig3}
Phase-folded radial-velocity measurements obtained with {\footnotesize
  CORALIE} for {\footnotesize HD}\,121504, {\footnotesize HD}\,142415,
{\footnotesize HD}\,147513, {\footnotesize HD}\,196050, {\footnotesize
  HD}\,216437 and {\footnotesize HD}\,216437, superimposed on the best
Keplerian planetary solution (top panel in each diagram). The
residuals as a function of Julian Date are displayed in the lower
panels.}
\end{center}
\end{figure*}

\subsection{The single-planet systems}

\begin{figure}[t!]
\begin{center}
\psfig{width=0.9\hsize,file=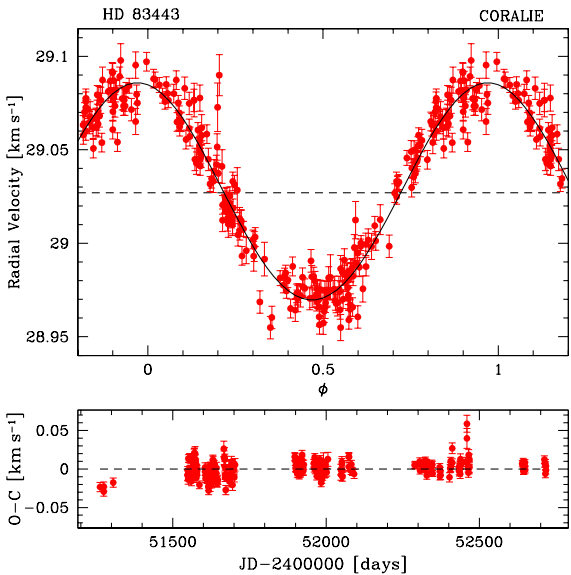}
\caption{
\label{fig4}
Phase-folded radial-velocity measurements obtained with {\footnotesize
  CORALIE} for {\footnotesize HD}\,83443, superimposed to the best
Keplerian planetary solution (top). The residuals as a function of
Julian Date are displayed in the lower panel.}
\end{center}
\end{figure}

To simplify the presentation of our results for the single-planet
systems, the standard orbital parameters derived from the best fitted
1-Keplerian solution to the data are gathered in Table\,\ref{table2}.
Useful inferred planetary parameters and further information on the
number of measurements used, their time coverage ($Span$) and the
residuals (weighted r.m.s.) around the solutions ($\sigma(O-C)$) are
given as well.  Finally, we also indicate the references for the
detection announcements of these candidates\footnote{At the IAU
  Symp.~202 in Manchester, we announced 6 {\small CORALIE} candidates
  \citep[][ 5 of them being described in this
  paper]{Mayor-2000:b,Queloz-2000:c,Udry-2000:b} and 1 further {\small
    ELODIE} candidate \citep{Sivan-2000}.  The proceedings of the IAU
  Symp.~202 have not yet appeared. Our 8 contributions to this
  conference describing the planet and brown-dwarf CORALIE and ELODIE
  new detections, the metallicity of stars hosting planets, the effect
  of stellar activity on radial-velocity measurements and a
  presentation of the new HARPS spectrograph (now running on the ESO
  3.6-m at La\,Silla) are accessible from our {\sl Exoplanets} web
  page: http://obswww.unige.ch/Exoplanets/publications.html}
($Det.Ref.$).  The corresponding phase-folded radial-velocity curves
are displayed in Figs.~\ref{fig2} to \ref{fig4} (top panel in each
diagram).  The residuals drawn as a function of the Julian date are
displayed in the figures as well (bottom panels).

Simultaneous independent discoveries have been made for 3 of our
candidates: by \citet{Fischer-2001} for {\footnotesize HD}\,92788 and
by \citet{Jones-2002} for {\footnotesize HD}\,196050 and
{\footnotesize HD}\,216437. Their orbital solutions are very similar
to ours (Table\,\ref{table2}).

For most of the systems, the residuals measured around the derived
solutions are compatible with the typical individual radial-velocity
uncertainties. Some of them are, however, slightly larger
($\sim$\,10\,m\,s$^{-1}$). In these cases, a slight level of activity
can be invoked to explain the additional noise. {\footnotesize
  HD}\,114386 is a good example with its clear activity level
testified by the visible reemission in the \ion{Ca}{ii}\,H absorption
line (Fig.\,\ref{fig1}).  Another illustration is given by the similar
early G dwarfs {\footnotesize HD}\,121504 and {\footnotesize
  HD}\,142415.  They present light rotation and an activity level that
can explain the somewhat large measured residuals.  It is interesting
to note here the similar values of $v\sin{i}$, $\log R^\prime_{HK}$
and $\sigma(O-C)$ for the two candidates.

In order to emphasize a relation between the residuals to the
Keplerian solutions and stellar activity, we compared in a systematic
way for all the candidates the obtained residuals with the shape of
the spectral lines, estimated by the bisector inverse slope of our
cross-correlation functions \citep[BIS;][]{Queloz-2001:a}.
Unfortunately, at this level of velocity variation
($\leq$\,10\,m\,s$^{-1}$) and at the S/N of the {\footnotesize
  CORALIE} spectra, no definitive conclusion can be drawn. The same
non-conclusive result was also obtained for {\footnotesize HD}\,73256
which did not show any clear relation between the BIS parameter and
the residuals around the 2.5-d solution, although the activity-induced
radial-velocity jitter was undubitably emphasized by the simultaneous
variations of the stellar photometric signal and the residuals
\citep{Udry-2003:c}. We also searched in a systematic way, through
Fourier analysis, for additional periodicities in the residuals around
the derived solutions. Nothing significant was found for these
candidates.  Some of them deserve, however, further comments (see the
following subsections).

Amongst the 13 single-planet systems presented in this subsection, 10
({\footnotesize HD}\,19994, {\footnotesize HD}\,65216, {\footnotesize
  HD}\,92788, {\footnotesize HD}\,111232, {\footnotesize HD}\,114386,
{\footnotesize HD}\,142415, {\footnotesize HD}\,147513, {\footnotesize
  HD}\,196050, {\footnotesize HD}\,216437, {\footnotesize HD}\,216770)
are on intermediate or long-period orbits
(100\,$<$\,$P$\,$\leq$\,1350\,days) with medium eccentricities
(0.2\,$\leq$\,$e$\,$\leq$\,0.5). Such properties are shared by the
bulk of the presently known exoplanets\footnote{Long-period candidates
  on low-eccentricity orbits (closer to the giant planets in our solar
  systems) are however more and more often discovered as the time
  coverage of the radial-velocity surveys increases}.  The
remaining 3 shorter-period planets ({\footnotesize HD}\,6434,
{\footnotesize HD}\,83443, {\footnotesize HD}\,121504) present, on the
other hand, a lower eccentricity. The eccentricity-period trend is
also observed for stellar binaries. This similitude between the two
populations was often brought up as a question for different planet
and binary formations. Nevertheless, clear differences exist
\citep[see e.g.][ for a more complete
discussion]{Mayor-2000:a,Halbwachs-2003}.  Also the mass-period
relation of exoplanets, checked for statistical significance by
\citet{Zucker-2002} and further discussed in \citet{Udry-2003:a}, is
apparent in our subsample; the two shortest-period systems host (by
far) the lightest planets, with minimum masses below 0.4\,M$_{\rm
  Jup}$, whereas the longest periods ($P$\,$\ge$\,1000\,d) are
associated with planet minimum masses of 1.82, 3.02 and 6.8\,M$_{\rm
  Jup}$.

\subsubsection*{Comments on HD\,6434}

Since the discovery of the planet, we have doubled the number of
measurements available for {\footnotesize HD}\,6434, gathering a total
of 130 good spectra over more than 1500~days.  The star relative
faintness and low metallicity lead to a typical radial-velocity
photon-noise uncertainty on individual measurements of only
$\sim$\,8\,m\,s$^{-1}$.  However, the large number of measurements
allows us to derive precise Keplerian orbital elements for the system
(Table\,\ref{table2}).  The very low planet minimum mass inferred from
the orbital solution ($m_2\sin{i}$\,=\,0.39\,M$_{\rm Jup}$) gives us a
first example of a very light planet orbiting a deficient star.

Although the star is found to be neither active nor rapidly rotating,
a concern is brought by the not so different values of the rotational
period (18.6\,day) estimated from the activity indicator and the
orbital period (22\,day). The uncertainty on the orbital period is
very small, but the calibrated $P_{rot}$ carries an intrinsic
uncertainty.  The a posteriori verification that the shape of the
spectral lines, estimated by the bisector inverse slope of our
cross-correlation functions \citep[BIS;][]{Queloz-2001:a}, does not
vary with the radial velocity provides an indication against activity
to be the source of the observed radial-velocity variation.  The BIS
itself is even constant with a r.m.s. of 8.6\,m\,s$^{-1}$, at the
photon-noise level. Moreover, the radial-velocity variation is stable
over more than 68 cycles.  However, rough simulations by
\citet{Santos-2003:b} raised the possibility of radial-velocity
variations without noticeable change in the BIS for slowly rotating
stars. In such cases, only simultaneous velocity and photometric
measurements can trace the intrinsic origin of the radial-velocity
variations, as e.g. for {\footnotesize HD}\,192263
\citep{Santos-2003:b} or for the {\footnotesize HD}\,73256 residuals
\citep{Udry-2003:b}. No indication of a 22-d periodicity is present in
the Hipparcos photometric data for {\footnotesize HD}\,6434 and the
Geneva photometry finds the star stable at a 3\,mmag level.

\subsubsection*{Comments on HD\,19994}

The residuals around the Keplerian solution (8.1\,m\,s$^{-1}$) are
slightly larger than individual photon-noise errors (median at
6\,m\,s$^{-1}$). They seem mainly due to velocities around
JD\,=\,2\,451\,600 presenting a large dispersion. Taking into account
the fairly large rotation of the star and its binarity status, this
could possibly come from an enhanced stellar activity level at that
moment, although the BIS parameter does not show anything special at
that time.  Nothing else particular is visible in the residuals.

\subsubsection*{Comments on HD\,83443}

\begin{figure}[t!]
\begin{center}
\psfig{width=0.9\hsize,file=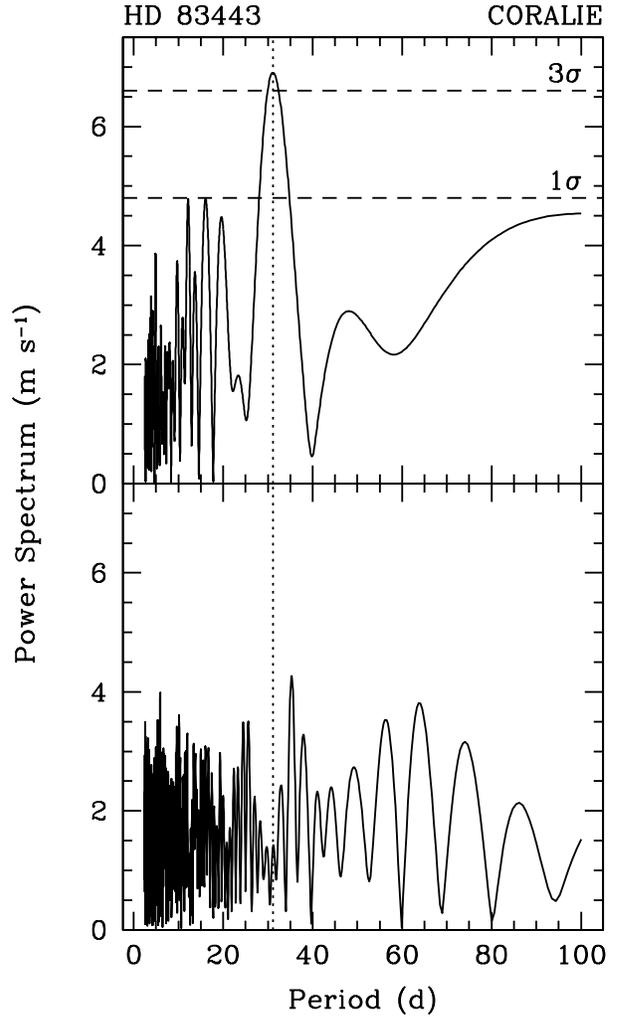}
\caption{
\label{fig5}
Fourier transform of the residuals around the 2.98-d orbital solution
of {\footnotesize HD}\,83443\,b {\sl Top:} for radial-velocities in
the 2\,451\,500\,$\leq$\,JD\,$\leq$\,2\,451\,702 period, when the 30-d
period signal was seen in the data, and {\sl Bottom:} for later
velocities. The false alarm probability of the 30-d period in the
early measurements is only $9\cdot 10^{-4}$ (close to the 4\,$\sigma$
detection level), whereas the signal has disappeared in the later
data.}
\end{center}
\end{figure}

{\footnotesize HD}\,83443 was first announced at the Manchester IAU
Symp.~202 to host a resonant 2-planet system with periods
$P_1$\,=\,2.985\,d and $P_2$\,=\,29.85\,d \citep{Mayor-2000:b}. At
that time the available 93 high-S/N {\footnotesize CORALIE} spectra
allowed us to clearly find the 2 periods of variation. The two
periodic signals were highly significant.  In particular, the false
alarm probability for the 30-d period in the data was very low
($\leq$\,$9\cdot 10^{-4}$), close to the $4\,\sigma$ detection limit
(Fig.\,\ref{fig5}, top).  Moreover, these periodicities were not
present in the velocities of contemporary observed constant stars with
similar photometric characteristics \citep{Udry-2002:b}. This rules
out instrumental effects as the source of the observed variations. A
new reduction of the data using the weighted cross-correlation scheme
also confirms the results presented in Manchester.

However, about 2 years later, \citet{Butler-2002} trying to confirm
the 2-planet system did not detect the {\nolinebreak 30-d}
second-planet signal in their Keck$+$AAT radial-velocity data.  We
then also verified that it had disappeared in our more recent data as
well \citep{Udry-2002:b}. The corresponding peak in the Fourier
transform of the radial velocities is no longer present
(Fig.\,\ref{fig5}, bottom).

The origin of this transient signal is not clear yet. An appealing
possibility is to attribute the effect to activity as the
corresponding period is compatible with the activity-related stellar
rotation estimate (within the uncertainties related to the
calibration). As mentioned above, we tried for this star as well to
see a relation between BIS and the residuals around the 2.98-d
Keplerian solution but without success at the precision of our
{\footnotesize CORALIE} data. No 30-d periodicity is found in the BIS
data either.  However, if existing, such a relation between residuals
and line shape should easily come out of the {\footnotesize HARPS}
data that are already providing a preliminary {\sl commissioning}
orbital solution for {\footnotesize HD}\,83443 at a
$\sim$\,1.5\,m\,s$^{-1}$ precision level \citep{Mayor-2003}.

Looking at the orbital solution given in Table\,\ref{table2}, we can
note that the ecccentricity derived for the planet is not
significantly different from 0. 

With its 2.98-day period, {\footnotesize HD}\,83443\,b was a good
candidate for photometric transit search.  The photometric $uvby$
observations were obtained with the {\sl Str\"omgren Automatic
  Telescope} ({\footnotesize SAT}) at {\footnotesize ESO} La\,Silla,
Chile (Olsen et al. in prep).  Unfortunately, no transit was detected.

\begin{table*}
\caption{
\label{table3}
{\footnotesize CORALIE} simulataneously derived 2-Keplerian orbital solutions 
as well as inferred planetary parameters for the multi-planet systems
{\footnotesize HD}\,82943 and {\footnotesize HD}\,169830. Parameter
definitions are the same as in Table\,\ref{table2}. Note that for
{\footnotesize HD}\,82943, the $e$ and $\omega$ elements are not well
defined in the sense that there exist other solutions with very different
$e$ and $\omega$ estimates but with equivalent $\sigma(O-C)$ values 
(see text).
}
\begin{tabular}{|l@{}l|r@{\,$\pm$\,}lr@{\,$\pm$\,}l|r@{\,$\pm$\,}lr@{\,$\pm$\,}l|}
\hline
  \multicolumn{2}{|l|}{\bf Parameter} 
& \multicolumn{2}{c}{\bf {\footnotesize HD}\,82943\,c} 
& \multicolumn{2}{c|}{\bf {\footnotesize HD}\,82943\,b} 
& \multicolumn{2}{c}{\bf {\footnotesize HD}\,169830\,b} 
& \multicolumn{2}{c|}{\bf {\footnotesize HD}\,169830\,c} \\
& & \multicolumn{4}{c|}{ ({\footnotesize HIP}\,47007)} 
&   \multicolumn{4}{c|}{ ({\footnotesize HIP}\,90485)} \\
$Det.Ref$. &      &\multicolumn{4}{c|}{\citet[May][]{ESO-2000};
\citet[April][]{ESO-2001}}  
                &\multicolumn{4}{c|}{\citet{Naef-2001,Udry-2003:e}} \\
\hline
$P$ &$[$days$]$     &219.4     &0.2    &435.1  &1.4   
                    &225.62    &0.22   &2102   &264 \\
$T$ &[JD-2\,400\,000]\hspace*{.2cm}   &52284   &1    &51758  &13    
                    &51923     &1      &52516  &25  \\ 
$e$ &               &0.38      &0.01   &0.18   &0.04  
                    &0.31      &0.01   &0.33   &0.02 \\   
$V$ &[km\,s$^{-1}$] &\multicolumn{4}{c|}{8.144\,$\pm$\,0.001} 
                    &\multicolumn{4}{c|}{$-17.209$\,$\pm$\,0.006} \\
$\omega$ &[deg]     &124       &3      &237    &13 
                    &148       &2      &252    &8  \\
$K$ &[m\,s$^{-1}$]  &61.5     &1.7    &45.8    &1.0  
                    &80.7     &0.9    &54.3    &3.6 \\
$a_1\sin i$ &[$\mathrm{10^{-3}}$\,AU]   
                &\multicolumn{2}{c}{1.145}  &\multicolumn{2}{c|}{1.801}  
                &\multicolumn{2}{c}{1.591}  &\multicolumn{2}{c|}{9.898}\\
$f(m)$ &$\mathrm{[10^{-9}\,M_{\odot}]}$ 
                &\multicolumn{2}{c}{4.156} &\multicolumn{2}{c|}{4.118} 
                &\multicolumn{2}{c}{10.56} &\multicolumn{2}{c|}{29.27} \\
$m_{2}\,\sin i$ &$\mathrm{[M_{\rm Jup}]}$ 
                &\multicolumn{2}{c}{1.85} &\multicolumn{2}{c|}{1.84}
                &\multicolumn{2}{c}{2.88} &\multicolumn{2}{c|}{4.04} \\
$a$ &[AU]       &\multicolumn{2}{c}{0.75} &\multicolumn{2}{c|}{1.18} 
                &\multicolumn{2}{c}{0.81} &\multicolumn{2}{c|}{3.60} \\
\hline
$N_{\rm meas}$ &   &\multicolumn{4}{c|}{142}  &\multicolumn{4}{c|}{112} \\
$Span$ &[days]     &\multicolumn{4}{c|}{1593} &\multicolumn{4}{c|}{1506}\\  
$\sigma (O-C)$\,\, &[m\,s$^{-1}$] 
                   &\multicolumn{4}{c|}{6.8}  &\multicolumn{4}{c|}{8.9}\\
$mask$ &           &\multicolumn{4}{c|}{weighted $K0$} 
                   &\multicolumn{4}{c|}{weighted $K0$} \\
\hline
\end{tabular}
\end{table*}

\subsubsection*{Comments on HD\,121504}

We mentioned above that the somewhat large residuals obtained for this
system may probably be related to the stellar activity-induced jitter
($\log{R^\prime_{HK}}$\,=\,$-4.57$). We also could see in the temporal
distribution of the residuals some indications of an additional
radial-velocity drift. If real, this drift is very small.  A combined
Keplerian~$+$~linear drift model yields a drift value of
$\sim$\,3\,m\,s$^{-1}$\,y$^{-1}$, without changing noticeably the
planetary orbital parameters. Furthermore, the combined fit does not
improve much the quality of the solution: the $\sigma(O-C)$ decreases
only from 11.6 to 11.2\,m\,s$^{-1}$. The drift is thus not considered
as significant and not included in the solution given in
Table\,\ref{table2}. Future measurements will confirm or rule out this
potential drift.

\begin{figure*}[t]
\begin{center}
\psfig{width=0.475\hsize,file=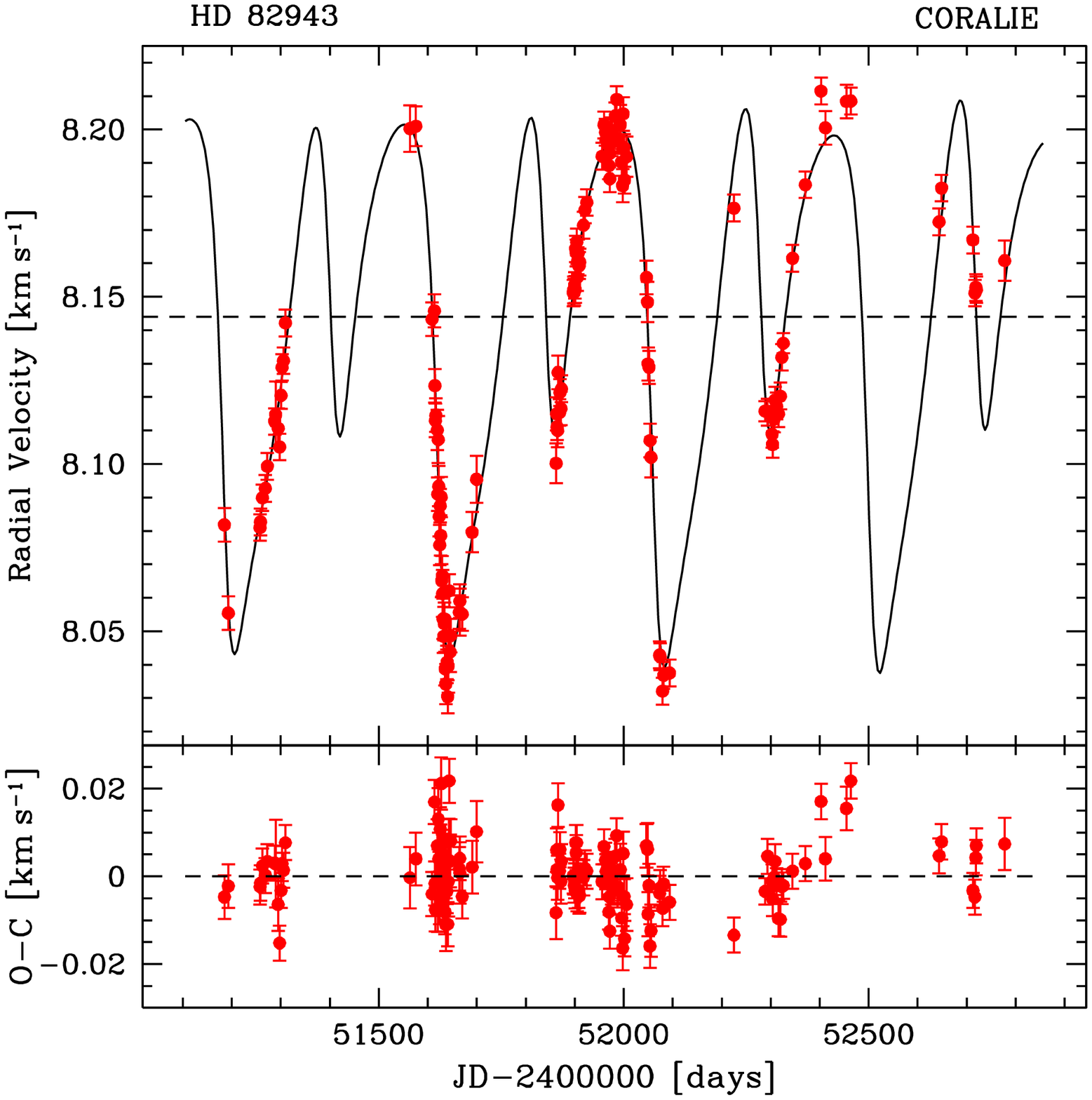}\,\,
\psfig{width=0.475\hsize,file=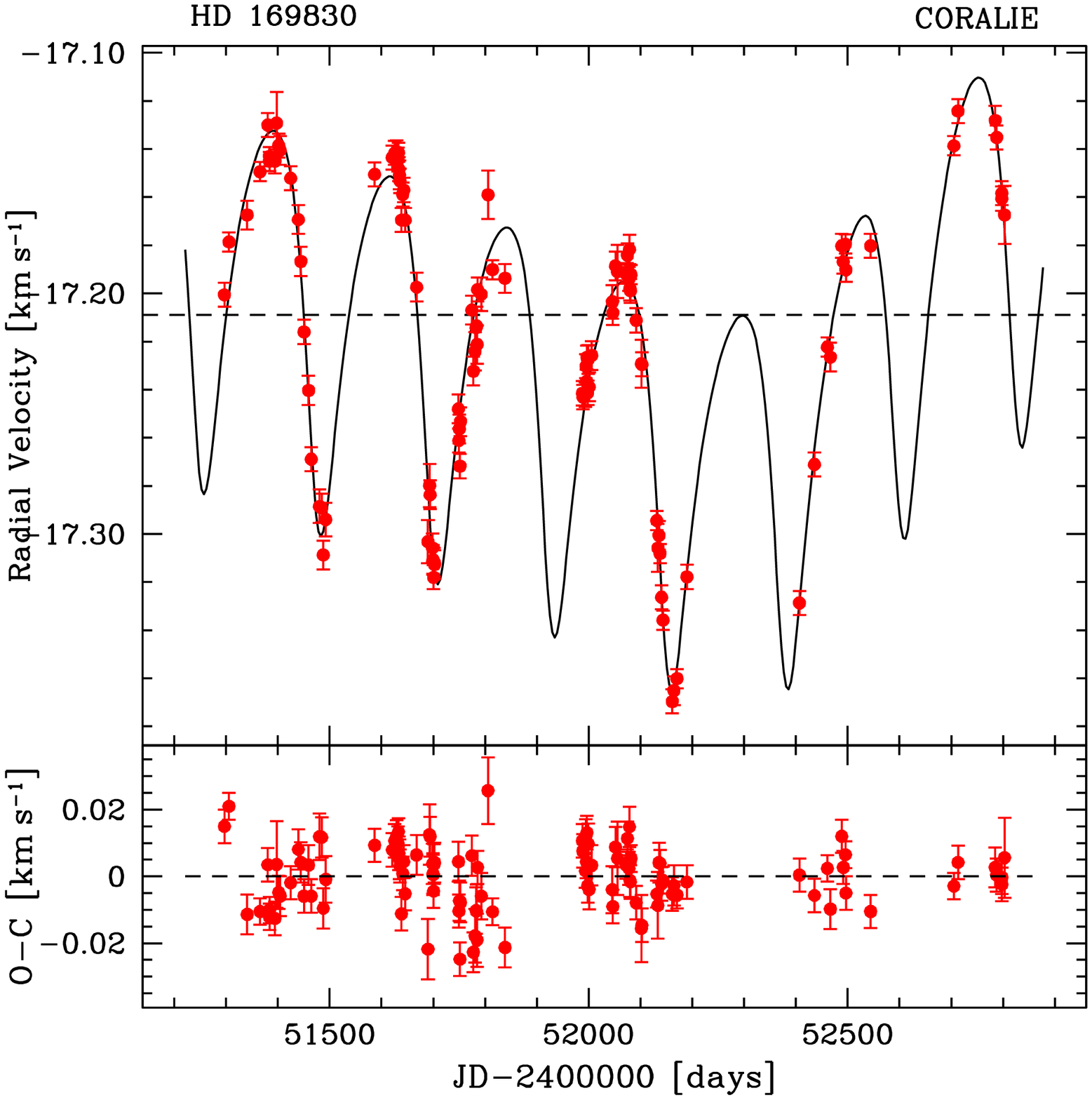}
\caption{
\label{fig6}
{\sl Top:} Temporal radial-velocity measurements obtained with
{\footnotesize CORALIE} for the 2-planet systems {\footnotesize
  HD}\,82943 (left) and {\footnotesize HD}\,169830 (right),
superimposed to the best simultaneously-derived 2-Keplerian planetary
solutions.  {\sl Bottom: } Residuals as a function of Julian Date.}
\end{center}
\end{figure*}

\subsubsection*{Comments on HD\,142415}

The Keplerian solution derived for {\footnotesize HD}\,142415 yields a
period of 386.3\,$\pm$\,1.6 days, not too far from 1 year. As we have
points covering the maximum and close (on both sides) to the minimum
of the phased radial-velocity curve over 4 full cycles, this proximity
to an annual variation is not so much of a concern for the reality of
the planetary signal. The trouble resides rather in the difficulty of
covering the whole phase interval.  As a result the minimum of the
curve is highly undersampled and the best fitted orbital solution
tends to be very eccentric, increasing the radial-velocity
semi-amplitude in an artificial way.  This effect is even worsened by
the stellar activity induced jitter that makes the orbital elements
more difficult to determine. We have thus arbitrarily fixed the
eccentricity of the solution to $e$\,=\,0.5, a plausible value when
looking at the sequence of the radial-velocity measurements. Note,
however, that solutions with $e$ between 0.2 and 0.8 would still be
acceptable. 

\subsubsection*{Comments on HD\,196050}

The phase coverage of our 31 {\footnotesize CORALIE} radial velocities
is not good enough to obtain a constrained value of the planet
eccentricity. We thus fixed this eccentricity to $e$\,=\,0.3, a value
minimizing the residuals around the derived solution.

\subsection{The 2-planet systems}

\subsubsection{The 2-planet resonant system around HD\,82943}

The star hosts a very interesting 2-planet system. Between
JD\,=\,2\,451\,184 and JD\,=\,2\,452\,777, we have gathered 142 good
{\footnotesize CORALIE} spectra of the star that allow us to derive a
2-Keplerian approximate solution with periods $P_b$\,=\,435.1\,d
  and $P_c$\,=\,219.4\,d.  The whole set of orbital parameters are
given in Table\,\ref{table3} and the solution and velocities are
displayed in Fig.\,\ref{fig6} (left).  The particular values of the
periods made us miss the short-period minimum of the curve when the
star was behind the Sun. The long-period planet ({\footnotesize
  HD}\,82943\,b) was thus announced first \citep{ESO-2000}, about 1
year before the detection of the secondary planet signature
\citep[{\footnotesize HD}\,82943\,c;][]{ESO-2001}.

The $P_i/P_j$\,=\,2/1 resonant systems are very important because,
when the planet orbital separations are not too large, planet-planet
gravitational interactions become non-negligible during planet
``close'' encounters, and will noticeably influence the system
evolution on a timescale of the order of a few times the long period.
The radial-velocity variations of the central star will then differ
substantially from velocity variations derived assuming the planets
are executing independent Keplerian motions. We observe a temporal
variation of the {\sl instantaneous} orbital elements. In the most
favourable cases, the orbital-plane inclinations, not otherwise known
from the radial-velocity technique, can be determined since the
amplitude of the planet-planet interaction directly scales with their
true masses.

In the case of multiple planets, only approximate analytic solutions
of the gravitational equations of motion exist, and one must resort to
numerical integrations to model the data.  Several studies have been
conducted in this direction for the {\footnotesize Gl}\,876 system
\citep{Laughlin-2001,Rivera-2001,Nauenberg-2002}, similar to
{\footnotesize HD}\,82943 in the sense that it also hosts two 2/1
resonant planets at fairly small separations. The results of the {\sl
  Newtonian} modeling of the {\footnotesize Gl}\,876 system have
validated the method, improving notably the determination of the
planetary orbital elements, but the time coverage of the measurements
is still too small for the method to provide strong constraints on the
plane inclinations. The valley of the acceptable solutions is still
very shallow, although including the correct answer provided by {\it
  HST} astrometric observations \citep{Benedict-2002}.  Further
radial-velocity measurements will undoubtedly improve the situation.

Our correct modeling of {\footnotesize HD}\,82943, taking into account
the planet-planet interactions, is under study and will be presented
in a forthcoming paper with a more detailed description of the system
behaviour (Correia et al. in prep).

Our present approximate solution for the system (Table\,\ref{table3})
yields residuals at the level of the photon noise of the
radial-velocity measurements. The inferred two planetary masses are
very similar.  Although in a 2/1 resonance, the two orbits do not seem
to be aligned.  However, we have to warn the reader not to take too
litterally the results described here because:

-- As mentioned above, due to the planet-planet interactions, the
orbital elements are time dependent;

-- Due to the non-optimum phase coverage, the planet eccentricities
and $\omega$s are badly (if at all) constrained by the data. The
small uncertainties given in the table only relate to the given local
solution in parameter space. There are, however, other local solutions
with almost equivalent $\chi^2$ minimum values. For instance, a
solution with the very different values of the eccentricities
$e_c$\,=\,0.4 and $e_b$\,$\simeq$\,0 only increases the residuals from
6.8 to 6.9\,m\,s$^{-1}$. In this case, as $e_b$\,$\simeq$\,0,
$\omega_b$ is completely non-determined and it is then possible to
find for these eccentricities an aligned configuration with
$\omega_b$\,=\,$\omega_c$\,=\,110\,deg and the same level of residuals
(6.9\,m\,s$^{-1}$), leaving furthermore $P_i$ and $K_i$ almost
unchanged.

We will now continue to follow closely this system, accumulating
measurements to improve in the future the derived solution.

\subsubsection{The HD\,169830 hierarchical 2-planet system}

A 230-d period planet orbiting {\footnotesize HD}\,169830 was first
described in \citet{Naef-2001}. The orbital solution was derived from
the 35 {\footnotesize CORALIE} spectra available at that time.  After
the second maximum of the radial-velocity curve, we noticed an
additional trend in the data pushing us to follow the star more
closely. We have now gathered 112 good spectra that allow us to
simultaneously derive a complete 2-Keplerian orbital solution for the
system (Table\,\ref{table3} and Fig.\,\ref{fig6}, right).

Unlike {\footnotesize HD}\,82943, this system is not resonant but more
hierarchically structured. The separation between the 2 planets stays
always fairly large, the 2-Keplerian model is then supposed to provide
a good approximation of the system evolution on a fairly long time.
However, the time span of our velocity measurements barely covers the
long period variation. The corresponding planetary orbit is thus not
completely constrained and a large uncertainty is still observed for
the long period.

This 2-planet system is the first to be announced after the
proposition by \citet{Mazeh-2003:a} of a possible correlation between
mass ratio and period ratio for adjacent planets in multi-planet
systems. It is interesting to note that the new system agrees with the
proposed correlation \citep{Mazeh-2003:b}.

\section{Summary}

We have described in this paper 16 still unpublished exo\-planet
candidates discovered with the {\footnotesize CORALIE} echelle
spectrograph mounted on the 1.2-m Euler Swiss telescope at La Silla
Observatory.  Amongst these new candidates:

-- Ten are typical extrasolar Jupiter-like planets on intermediate- or
long-period (100\,$<$\,$P$\,$\leq$\,1350\,d) and fairly eccentric
(0.2\,$\leq$\,$e$\,$\leq$\,0.5) orbits ({\footnotesize HD}\,19994,
{\footnotesize HD}\,65216, {\footnotesize HD}\,92788, {\footnotesize
  HD\,111232}, {\footnotesize HD}\,114386, {\footnotesize HD}\,142415,
{\footnotesize HD}\,147513, {\footnotesize HD}\,196050, {\footnotesize
  HD}\,216437, {\footnotesize HD}\,216770). They resemble the bulk of
extra-solar planets found to date.

-- Two of these planets ({\footnotesize HD}\,19994, {\footnotesize
  HD}\,147513) are orbiting one component of a multiple-star system.
Such planets seem to present different orbital and mass
characteristics than the other {\sl single}-star planets
\citep{Zucker-2002, Eggenberger-2003}. The companion to {\footnotesize
  HD}\,147513 is even a white dwarf, the evolution to which has
probably also influenced the planet evolution through mass transfer
between the two stars.

-- Three candidates are shorter-period planets ({\footnotesize
  HD}\,6434, {\footnotesize HD}\,121504, {\footnotesize HD}\,83443)
with lower eccentricities (the latter being a hot Jupiter).  

-- More interesting cases are given by the multiple-planet systems
{\footnotesize HD}\,82943 and {\footnotesize HD}\,169830.
{\footnotesize HD}\,82943 is a resonant $P_b/P_c$\,=\,2/1 system in
which planet-planet interactions are influencing the system evolution.
{\footnotesize HD}\,169830 is non-resonant and more hierarchically
structured, and therefore less affected by this kind of interaction.

From a more global point of view, our candidates follow the
period-eccentricity and period-mass trends observed for the whole
sample of known extra-solar planets.  They follow as well the trend
for stars hosting planets to be more metal rich than {\sl normal}
stars of the solar neighbourhood
\citep{Santos-2001:a,Santos-2003:a,Gonzalez-2001,Laws-2003}. Only 3
amongst the 15 stars are metal deficient with regards to the Sun
whereas almost all the others present high [Fe/H] values.

We emphasize the difficulty encountered to fully constrain
multi-planet systems. Such a task, involving many free parameters,
requires a good phase coverage and a fair number of measurements, even
for the simplest cases.  As a consequence, studies on multi-planet
system stability should not rely too closely on the given orbital
parameters.  The published solutions will probably change in the
future (some will notably change) as more measurements become
available. A substantial advance in this domain will be brought by the
the new {\footnotesize HARPS} spectrograph mounted on the
{\footnotesize ESO} 3.6-m telescope at La~Silla \citep{pepe-2002:b}
available since October 2003.  With the very high precision achieved
for radial-velocity measurements and the quality of the spectra,
{\footnotesize HARPS} is now providing us with an unequalled tool to
characterize multi-planet systems and/or disentangle activity-induced
jitter from orbital radial-velocity variations.


\begin{acknowledgements}
  We are grateful to the staff from the Geneva Observatory which is
  maintaining the 1.2-m Euler Swiss telescope and the CORALIE echelle
  spectrograph at La\,Silla. In particular, many thanks to Luc Weber
  for his continuous improvement of the {\footnotesize CORALIE}
  spectrograph softwares and to Bernard Pernier for his efforts in
  maintaining the {\footnotesize CORALIE} database and for his
  contribution to a large number of observations.  We thank the Geneva
  University and the Swiss NSF (FNRS) for their continuous support for
  this project.  Support from Funda\c{c}\~ao para a Ci\^encia e
  Tecnologia, Portugal, to N.C.S., in the form of a scholarship is
  gratefully acknowledged.  This research has made use of the Simbad
  database, operated at CDS, Strasbourg, France
\end{acknowledgements}


\bibliographystyle{aa} 
\bibliography{udry_articles}

\end{document}